\documentclass[12pt]{iopart}
\usepackage[dvipdfmx]{graphicx}
\usepackage{amsmath}
\usepackage{amssymb}
\usepackage{amscd}
\usepackage{bm}

\usepackage[mathscr]{eucal}

\newtheorem{theorem}{Theorem}
\newtheorem{lemma}[theorem]{Lemma}

\newtheorem{proposition}[theorem]{Proposition}
\newtheorem{corollary}[theorem]{Corollary}

\newtheorem{example}[theorem]{Example}

\newtheorem{remark}[theorem]{Remark}

\newcommand{\tp}[1]{\mathsf{Trop}({#1})}

\newcommand{\Z}{{\mathbb Z}}
\newcommand{\R}{{\mathbb R}}
\newcommand{\C}{{\mathbb C}}

\newcommand{\wt}{{\rm wt}}

\newcommand{\proof}{\noindent \textit{Proof. }}
\newcommand{\qed}{\hfill $\Box$}
\newcommand{\ve}{\varepsilon}

\newcommand{\batten}[4]{%
\begin{picture}(40,40)(-20,-20)
	\put(-10,0){\line(1,0){20}}
	\thicklines
	\put(0,10){\line(0,-1){20}}
	\put(-11,0){\makebox(0,0)[r]{$#1$}}
	\put(0,11){\makebox(0,0)[b]{$#2$}}
	\put(0,-11){\makebox(0,0)[t]{$#3$}}
	\put(11,0){\makebox(0,0)[l]{$#4$}}
\end{picture}
}

\begin{document}
\title{Geometric lifting of the integrable cellular automata with periodic boundary conditions}

\author{Taichiro Takagi and Takuma Yoshikawa}
\address{Department of Applied Physics, National Defense Academy, Yokosuka, Kanagawa 239-8686, Japan}
\ead{takagi@nda.ac.jp \textrm{and} em58038@nda.ac.jp}
\begin{abstract}
Inspired by G. Frieden's recent work on the geometric $R$-matrix for
affine type $A$ crystal associated with rectangular shaped Young tableaux,
we propose  a method to
construct a novel family of discrete integrable systems 
which can be regarded as
a geometric lifting of the generalized periodic box-ball systems.
By converting the conventional usage of the 
matrices for defining the Lax representation of the
discrete periodic Toda chain,
together with a clever use of the Perron-Frobenious theorem,
we give a definition of our systems.
It is carried out
on the space of real positive dependent variables,
without regarding them to be written by subtraction-free rational functions of
independent variables
but nevertheless with the conserved quantities which can be tropicalized.
We prove that, in this setup an equation of an analogue of the `carrier'
of the box-ball system for
assuring its periodic boundary condition always has a unique solution.
As a result, any states in our systems admit a commuting family of time evolutions
associated with any rectangular shaped tableaux,
in contrast to the case of corresponding
generalized periodic box-ball systems
where some states did not admit some of such time evolutions.
\end{abstract}
\title[Geometric lifting of the integrable cellular automata]
\maketitle

\section{Introduction}\label{sec:1}
\subsection{Backgrounds and main results}\label{sec:1_1}
Integrable systems in classical and quantum theory
have attracted many attentions from those studying
in the field of
mathematical physics.
Related to both classical and quantum integrable systems,
the integrable cellular automata known as the box-ball systems
have provided many stimulating ideas and topics 
in this field \cite{IKT12, TS90}.
In relation to classical integrable systems, the box-ball systems are derived from
integrable non-linear differential equations by a procedure
known as tropicalization or ultra-discretization. 
Roughly speaking, the notion of geometric lifting is 
the inverse of this procedure.

On the other hand,
in relation to quantum integrable systems, the box-ball systems are derived from
integrable quantum spin chain models by a procedure crystallization.
Its mathematical background is given by Kashiwara's theory of crystals \cite{Ka1, Ka2}.
Quite remarkably, there is a geometric lifting of this theory
known as the theory of geometric (and unipotent) crystals by
Berenstein and Kazhdan \cite{BK00}.
Based on their work,
G.~Frieden recently presented
explicit formulas for the 
affine type $A$ geometric crystal and its intertwiner, the
geometric $R$-matrix, 
by using Grassmannians \cite{F19, F18}.
This is a geometric lifting of the crystal of the so-called 
Kirillov-Reshetikhin modules and the associated
combinatorial $R$-matrix,
represented by semi-standard Young tableaux with
rectangular shapes. 

Inspired by his work, in this paper we propose a method to
construct
a geometric lifting of a family of integrable cellular automata
with periodic boundary conditions.
These cellular automata are known as the
periodic box-ball system \cite{YYT, YT02} and
its generalizations \cite{KS08, KTT, KT10, KT2}.
Here we note that the periodic box-ball system was conventionally derived from
time-discretized version of the closed Toda chain
(or discrete periodic Toda; dp-Toda \cite{HT95, HTI93})
by the
procedure tropicalization \cite{KT02, IT08, T14}.
Therefore one may think that its geometric lifting simply goes back to the
original dp-Toda chain.
However,  our method of geometric lifting is considerably different from that, and
gives a novel family of discrete integrable systems,
which we call \textit{closed geometric crystal chains}.

In order to explain the difference between the dp-Toda chain and 
the closed geometric crystal chain, we use the notion of
discrete time Lax equation \cite{Suris04}.
Let $\mathbf{x} = (x^{(1)},\dots,x^{(n-1)})$ be an $(n-1)$-component variable
and set $x^{(n)}:=s/(x^{(1)} \cdots x^{(n-1)})$
where $s$ is a parameter in $\C^{\times}$ or $\R_{>0}$.
We introduce an $n \times n$ matrix $g(\mathbf{x}, s; \lambda)$ as in the 
main text (See Example \ref{ex:nov20_2} for $n=4$), 
in which the diagonal elements are $(x^{(1)},\dots,x^{(n)})$,
their nearest lower off-diagonals are $1$'s, and there is
an indeterminate $\lambda$ at the top-right corner.
Then, for any $s,l \in \C^{\times}$ and sufficiently generic
$(\mathbf{a},\mathbf{b}) \in (\C^{\times})^{2(n-1)}$,
there is a unique solution $(\mathbf{a}',\mathbf{b}') \in (\C^{\times})^{2(n-1)}$
to the following matrix equation
\begin{equation}\label{eq:dec11_5}
g(\mathbf{b}, s; \lambda) g(\mathbf{a}, l; \lambda) =
g(\mathbf{a}', l; \lambda) g(\mathbf{b}', s; \lambda).
\end{equation}
By regarding the map $T: (\mathbf{a},\mathbf{b}) \mapsto (\mathbf{a}',\mathbf{b}')$
as a time evolution, we obtain a non-linear dynamical system on 
$ (\C^{\times})^{2(n-1)}$ which 
(with a shift of the indices of the variables' components)
turns out to be the dp-Toda chain.
Its Lax representation is given by
\begin{equation*}
\mathcal{L} (\mathbf{a}',\mathbf{b}' ; \lambda) =
(\mathcal{M}(\lambda))^{-1} \mathcal{L} (\mathbf{a},\mathbf{b} ; \lambda)
\mathcal{M}(\lambda),
\end{equation*}
where $\mathcal{L} (\mathbf{a},\mathbf{b} ; \lambda) = 
g(\mathbf{a}, l; \lambda) g(\mathbf{b}, s; \lambda)$ and
$\mathcal{M}(\lambda) = (g(\mathbf{b}, s; \lambda))^{-1}$.
We note that this map is an example of what are called
integrable or Yang-Baxter maps \cite{Sklyanin00, V05}.

For the dp-Toda chain, it is known that
every component of the dependent variables $(\mathbf{a}',\mathbf{b}')$ is
expressed by a subtraction-free rational function of the parameters
$s,l$ and the components of the independent variables $(\mathbf{a},\mathbf{b})$
with non-negative integer coefficients.
This implies that, if we let the parameters $s,l$ to take their values in $\R_{>0}$, 
then we can regard  
the dp-Toda chain as a dynamical system on 
$ (\R_{>0})^{2(n-1)}$ instead of $(\C^{\times})^{2(n-1)}$.
This idea of restricting the domains of the parameters and the
variables into \textit{positive real} spaces
opens a door to the possibility of constructing a new family of
discrete integrable systems out of this well-known matrix $g(\mathbf{x}, s; \lambda)$.
More precisely, we are going to adopt a new guiding principle of our study
for seeking such integrable systems that have positive real dependent variables
but now without regarding them to be written by subtraction-free rational
functions of independent variables.

%
To be more explicit, 
one of the main results of this paper (Theorem \ref{th:main}) claims that
for any $L \in \Z_{>0}$,
$s,l \in \R_{>0}$ and $(\mathbf{b}_1, \dots, \mathbf{b}_L) 
\in  (\R_{>0})^{L(n-1)} $,
there is a unique \textit{positive real solution} 
$(\mathbf{v}, \mathbf{b}'_1, \dots, \mathbf{b}'_L) \in(\R_{>0})^{(L+1)(n-1)}$ to the 
following matrix equation
\begin{equation}\label{eq:dec10_1}
g(\mathbf{b}_1, s; \lambda) \cdots g(\mathbf{b}_L, s; \lambda) g(\mathbf{v}, l; \lambda) =
g(\mathbf{v}, l; \lambda) g(\mathbf{b}'_1, s; \lambda) \cdots g(\mathbf{b}'_L, s; \lambda).
\end{equation}
This remarkably simple result is obtained by combining Frieden's work on the 
geometric $R$-matrix \cite{F18} with the Perron-Frobenius theorem
in linear algebra (See, for example \cite{Lax96}).
Therefore, by regarding the map $T^{(1)}_l: (\mathbf{b}_1, \dots, \mathbf{b}_L) 
\mapsto (\mathbf{b}'_1, \dots, \mathbf{b}'_L)$
as a time evolution, we obtain a non-linear dynamical system on $ (\R_{>0})^{L(n-1)} $. 
This is an example of the closed geometric crystal chains.
Obviously, its Lax representation is given by
\begin{equation*}
\mathcal{L} (\mathbf{b}'_1, \dots, \mathbf{b}'_L ; \lambda) =
(\mathcal{M}(\lambda))^{-1} \mathcal{L} (\mathbf{b}_1, \dots, \mathbf{b}_L ; \lambda)
\mathcal{M}(\lambda),
\end{equation*}
where $\mathcal{L} (\mathbf{b}_1, \dots, \mathbf{b}_L ; \lambda) = 
g(\mathbf{b}_1, s; \lambda) \cdots g(\mathbf{b}_L, s; \lambda)$ and
$\mathcal{M}(\lambda) = g(\mathbf{v}, l; \lambda)$.
Although we adopted the above mentioned guiding principle,
this Lax representation allows us to obtain such conserved quantities that
still can be tropicalized.

Now we want to explain how the above defined closed geometric crystal chains
are related to integrable cellular automata with periodic boundary conditions.
The geometric $R$-matrix (in the totally one-row tableaux case)
is a map defined to be
$R: (\mathbf{b},\mathbf{a}) \mapsto (\mathbf{a}',\mathbf{b}')$
in which the variables are related by
the matrix equation \eqref{eq:dec11_5}.
For a state $(\mathbf{b}_1, \dots, \mathbf{b}_L) 
\in  (\R_{>0})^{L(n-1)} $ and a `carrier' $\mathbf{v} \in  (\R_{>0})^{n-1}$
we use this map repeatedly.
It is illustrated as
\begin{equation}\label{eq:dec11_6}
\batten{\mathbf{v}_1}{\mathbf{b}_1}{\mathbf{b}_1'}{\mathbf{v}_2}\!\!\!
\batten{}{\mathbf{b}_2}{\mathbf{b}_2'}{\mathbf{v}_3}\!\!\!
\batten{}{}{}{\cdots\cdots}
\quad
\batten{}{}{}{\mathbf{v}_{L-1}}\,\,
\batten{}{\mathbf{b}_{L-1}}{\mathbf{b}_{L-1}'}{\mathbf{v}_{L}}\!\!\!
\batten{}{\mathbf{b}_L}{\mathbf{b}_L'}{\mathbf{v}.}
\end{equation}
We warn that this diagram should be read from the right to the left.
So the variables are related by 
$R(\mathbf{b}_i,\mathbf{v}_{i+1})=(\mathbf{v}_{i}, \mathbf{b}'_i)$
where we interpret $\mathbf{v}_{L+1}$ as $\mathbf{v}$.
In the corresponding combinatorial theory,
the geometric $R$-matrix is tropicalized to
the combinatorial $R$-matrix.
This is a map for the isomorphism of the tensor products
of Kashiwara's crystals.
As a combinatorial analogue of the relation
depicted by \eqref{eq:dec11_6},
we present an example in the case of $n=2$ cited from
\cite{KTT}.
\begin{equation*}
\unitlength 0.8mm
{\small
\begin{picture}(130,20)(8,-11)
\multiput(0.8,0)(11.5,0){13}{\line(1,0){4}}
\multiput(2.8,-3)(11.5,0){13}{\line(0,1){6}}

\put(1.8,5){2}
\put(13.3,5){1}
\put(24.8,5){2}
\put(36.3,5){2}
\put(47.8,5){1}
\put(59.3,5){1}
\put(70.8,5){1}
\put(82.3,5){1}
\put(93.8,5){2}
\put(105.3,5){2}
\put(116.8,5){2}
\put(128.3,5){1}
\put(139.8,5){1}

\put(1.8,-7.5){1}
\put(13.3,-7.5){2}
\put(24.8,-7.5){1}
\put(36.3,-7.5){1}
\put(47.8,-7.5){1}
\put(59.3,-7.5){2}
\put(70.8,-7.5){2}
\put(82.3,-7.5){2}
\put(93.8,-7.5){1}
\put(105.3,-7.5){1}
\put(116.8,-7.5){1}
\put(128.3,-7.5){2}
\put(139.8,-7.5){2}

\put(-6,-1.1){122}
\put(5.5,-1.1){112}
\put(17,-1.1){122}
\put(28.5,-1.1){112}
\put(40,-1.1){111}
\put(51.5,-1.1){111}
\put(63,-1.1){112}
\put(74.5,-1.1){122}
\put(86,-1.1){222}
\put(97.5,-1.1){122}
\put(109,-1.1){112}
\put(120.5,-1.1){111}
\put(132,-1.1){112}
\put(143.5,-1.1){122}
\end{picture}
}
\end{equation*}
(This is a `reflected' version of \cite{KTT} 
where the left and right
hand sides have been inverted.)
In the language of the box-ball systems,
the single letters $1$ and $2$ denote
an empty box and a box with a ball respectively,
where the capacity of the boxes are all one.
The three consecutive letters' $111, 112, \dots$
on the middle horizontal line denote the states of a carrier of
balls with capacity three, that travels from the right to the left.
The carrier picks up a ball from the box with a ball or
put a ball into an empty box, if possible in either case,
at each site on the way of the traveling.
In \cite{KTT}, A.~Kuniba, A.~Takenouch and one of the authors
proved that:
\begin{proposition}\label{pr:dec11_7}
Suppose $n=2$ and the state is given by a sequence of 
single box tableaux.
Then for any capacity of the carrier,
tropical analogue of the equation $\mathbf{v}_1 = \mathbf{v}$
for the picture \eqref{eq:dec11_6} has
at least one solution,
and that even if there are more than one solution to this equation,
tropical analogue of the state $\mathbf{b}'_1, \dots, \mathbf{b}'_L$
on the bottom line is independent of the choice of the 
non-unique solutions and hence is
uniquely determined.
\end{proposition}

This fact enabled us to 
give a formulation of the periodic box-ball system
in terms of Kashiwara's crystal theory,
and it is natural to consider a generalization of
this formulation to those associated with crystals of
Kirillov-Reshetikhin modules.
In this generalization, the combinatorial analogue of the relation
$R(\mathbf{b}_i,\mathbf{v}_{i+1})=(\mathbf{v}_{i}, \mathbf{b}'_i)$
is given by a relation between \textit{product tableaux} \cite{Ful}.
That is, it is given by
$\tp{\mathbf{b}_i} \cdot \tp{\mathbf{v}_{i+1}} =
\tp{\mathbf{v}_i} \cdot \tp{\mathbf{b}'_{i}}$ where
$\tp{\bullet}$ denotes a rectangular tableau with $k$ rows for $1 \leq k \leq n-1$
obtained by the tropicalization of 
any element of $(\R_{>0})^{k(n-k)}$.
Here is an example cited from \cite{KT10}
for $n=4$ where the carrier is given by a tableau with two rows.  
%
\begin{equation*}
\footnotesize
\begin{picture}(200,30)(30,-7)

\put(18,14){1}\put(48,14){3}\put(78,14){4}
\put(108,14){3}\put(138,14){2}\put(168,14){1}
\put(198,14){1}\put(228,14){4}\put(258,14){2}

\multiput(20,3)(30,0){9}{
\put(-6,0){\line(1,0){12}}\put(0,-8){\line(0,1){16}}}

\put(270,3.5){12}\put(270,-3.5){34}
\put(240,3.5){12}\put(240,-3.5){23}
\put(210,3.5){13}\put(210,-3.5){24}
\put(180,3.5){11}\put(180,-3.5){24}
\put(150,3.5){11}\put(150,-3.5){24}
\put(120,3.5){11}\put(120,-3.5){22}
\put(90,3.5){12}\put(90,-3.5){23}
\put(60,3.5){13}\put(60,-3.5){24}
\put(30,3.5){23}\put(30,-3.5){34}
\put(0,3.5){12}\put(0,-3.5){34}

\put(18,-14){3}\put(48,-14){1}\put(78,-14){2}
\put(108,-14){1}\put(138,-14){4}\put(168,-14){1}
\put(198,-14){3}\put(228,-14){2}\put(258,-14){4}
\end{picture}
\normalsize
\end{equation*}
In this example, the above product tableaux relation
can be described in such a way that
the column-insertion of $\tp{\mathbf{b}_i} $ into $\tp{\mathbf{v}_{i+1}} $
coincides with the row-insertion of 
$\tp{\mathbf{b}'_i} $ into $\tp{\mathbf{v}_{i}}$.
This example allures us to have a dream that
we might be able to construct associated integrable cellular automata,
in the sense that 
for any sequence of letters arbitrarily chosen
from the set $\{1,\dots,n\}$
and for any rectangular shape,
we can always find a tableau of that shape and
solves
tropical analogue of the equation $\mathbf{v}_1 = \mathbf{v}$
for the picture \eqref{eq:dec11_6},
allowing us to define a unique time evolution
compatible with the periodic boundary condition.
In fact, such a dream does not come true because the analogue of
Proposition \ref{pr:dec11_7} does not hold 
for $n>2$ \cite{KT10, KT2}, and even in the $n=2$ case it does not hold
for sequences of general one-row tableaux \cite{KS08, T09}.

The motivation for beginning our present study was
to clarify how this situation would be changed if the combinatorial $R$-matrices
are lifted to the geometric $R$-matrices.
The outcome is a realization of the above mentioned dream, in a sense.
The main result of this paper is Theorem \ref{th:main2}, that
generalizes the above mentioned Theorem \ref{th:main} from the 
totally one-row tableaux case to the case of carriers of
general rectangular tableaux.

\subsection{Outline}\label{sec:1_2}
Throughout this paper,
a notation for
a positive integer $n \geq 2$ is fixed which comes from
such usages as in the theory of
type $A^{(1)}_{n-1}$ geometric crystals, 
semi-standard Young tableaux with the entries taken from
$\{ 1, \dots , n\}$,
or the loop group ${\rm GL}_n (\C (\lambda))$.

In section \ref{sec:2}, we restrict ourselves to the simplest case of $n=2$
and give a detailed description of the simplest nontrivial example
of our new integrable systems, the closed geometric crystal chains.
In section \ref{sec:2_1}, by using only elementary mathematics
we show that the above mentioned scheme of
constructing a new dynamical system associated with the matrix 
$g(\mathbf{x}, s; \lambda)$
is indeed possible by the restriction of the variables
to the positive real domains. 
In section \ref{sec:2_2}, we study the properties of the
dynamical system and clarify its integrable structures such as
descriptions of its
conservation laws.
In section \ref{sec:2_3}, 
we consider two different kinds of continuum limits of
our discrete time dynamical system to derive its associated differential
equations in scope for potential application to real physical systems.
In section \ref{sec:2_4}, we study tropicalization of our dynamical system
to elucidate its relation to the generalized periodic box-ball systems.

Extension to the case of
general $n$ is explored in section \ref{sec:3}.
This section is divided into two subsections
according to the shapes of rectangular Young tableaux 
whose geometric/rational lifts are used there.
In section \ref{sec:3_1}, we use one-row tableaux only and
consider the matrix equation \eqref{eq:dec10_1} to
define time evolutions for our dynamical system, as well as to study 
its conservation laws.
In section \ref{sec:3_2}, we still use one-row tableaux only for
the \textit{states} of our dynamical system, but use general rectangular
tableaux for the \textit{carriers}, that would play the role of
carriers of balls in the associated box-ball systems with
$n-1$ species of balls.
To this end, we present a brief review on the 
geometric $R$-matrix introduced by Frieden, and
find a way to use its properties and the Perron-Frobenious theorem to
our construction of commuting time evolutions for the
new integrable systems.

Finally, in section \ref{sec:4} we give a summary and discussions.

\subsection{Notation}\label{sec:1_3}
As explained in the above,
we fix a notation $n \geq 2$ for an integer,
and we write $[n]=\{1,\dots,n\}$.
For any $r \in [n]$, denote
${ [n] \choose r }$ to be the set of $r$-element subsets of $[n]$.
For any admissible pair of $r$-element subsets $I, J$ and any matrix $A$ which has
more than or equal to $r$ rows and columns,
denote $\Delta_{I,J}(A)$ to be a minor determinant of $A$
associated with its $r \times r$ submatrix 
specified by rows in $I$ and columns in $J$.
For two integers $i$ and $j$, we write $[i,j]=\{m \in \Z \vert i \leq m \leq j\}$.

Denote ${\rm Gr}(r,n)$ to be the Grassmannian variety of $r$-dimensional
subspaces in $\C^n$.
For $J \in { [n] \choose r }$ we write $P_J(M)$ to denote the 
$J$th Pl$\ddot{\rm u}$cker coordinate of the subspace $M \in {\rm Gr}(r,n)$.  
Pl$\ddot{\rm u}$cker coordinates are projective, i.~e.~they are only defined
up to a common nonzero scalar multiple.

For the Pl$\ddot{\rm u}$cker coordinates,
in most cases we adopt Convention 3.1 of \cite{F19}.
We often write $P_1, P_{12}, P_{123}$ instead of $P_{\{1\}}, P_{\{1,2\}}, P_{\{1,2,3\}}$.
If $I \in [n]$ does not contain exactly $r$ elements, then we set
$P_I(M)=0$.
If $I$ is any set of integers, we set $P_I(M) = P_{I'}(M)$, where $I'$ is the set 
consisting of the residues of the elements of $I$ modulo $n$,
where the residues are thought of as elements of $[n]$.

A point $M \in {\rm Gr}(r,n)$ is represented by a full-rank $n \times r$
matrix $M'$, in the sense that its columns span the subspace $M$.
Thus, for any $B \in {\rm GL}_r(\C)$ the matrix $M' B$ represents
the same point $M$.
This enables us to write the (projective) Pl$\ddot{\rm u}$cker coordinate
as $P_J(M) = \Delta_{J, [r]} (M')$,
because we have $\Delta_{J, [r]} (M'B)= \Delta_{J, [r]} (M') \cdot \det B$
by the Cauchy-Binet formula.
In contrast, for any $A \in {\rm GL}_n(\C)$ the matrix $A M'$ represents
generally another point in ${\rm Gr}(r,n)$ that is denoted by $A \cdot M$.

We write $\mathbb{I}_r$ to denote the $r \times r$ identity matrix.

\section{The case of $\bm{n=2}$}\label{sec:2}
\subsection{Definition of the dynamical system}\label{sec:2_1}
We first consider the simplest case of the geometric $R$-matrix 
that is the geometric lifting of the combinatorial $R$-matrix
for 
one-row Young tableaux with $2$ kinds of letters.
Let $s,l \in \R_{>0}$ be a pair of parameters, and $R: (\R_{>0})^2 \rightarrow  (\R_{>0})^2$
a rational map given by $R:(b,a) \mapsto (a',b')$ where
\begin{equation}\label{eq:jul28_1}
a' = a \frac{b+\frac{l}{a}}{a+\frac{s}{b}}, \quad b' =b \frac{a+\frac{s}{b}}{b+\frac{l}{a}}.
\end{equation}
We depict the relation $R(b,a) = (a',b')$
by
\begin{equation*}
\batten{a'}{b}{b'}{a}.
\end{equation*}
If necessary, we denote by $R^{(s,l)}$ for $R$ to explicitly express 
its dependence on the parameters $s,l$.
It is easy to see that $R^{(l,s)} \circ R^{(s,l)}= {\rm Id}$, so in particular
this map is birational. 

Let $R_i$ be a map from $(\R_{>0})^{L+1}$ to itself,
which acts as the map $R$ on factors $i$ and $i+1$, and as the identity on
the other factors.
Let $\mathcal{R} = R_1 \circ \cdots \circ R_L$.
Given an arbitrary $(b_1, \dots, b_L, v) \in (\R_{>0})^{L+1}$
let $\mathcal{R} (b_1, \dots, b_L, v) = (v_1, b'_1, \dots, b'_L)$.
It is depicted by
\begin{equation}\label{eq:jul28_2}
\batten{v_1}{b_1}{b_1'}{v_2}\!\!\!
\batten{}{b_2}{b_2'}{v_3}\!\!\!
\batten{}{}{}{\cdots\cdots}
\quad
\batten{}{}{}{v_{L-1}}\,\,
\batten{}{b_{L-1}}{b_{L-1}'}{v_{L}}\!\!\!
\batten{}{b_L}{b_L'}{v.}
\end{equation}
Based on this diagram,
we would like to construct a
discrete time dynamical system on the space $(\R_{>0})^L$
using a map that sends $(b_1, \dots, b_L)$ to $(b'_1, \dots, b'_L)$
as a unit step of its time evolution.
Then, if the $v_1$ appeared at the left end coincides with $v$
at the right end, it is reasonable to say that this one-dimensional system is
satisfying a periodic boundary condition.
Note that
%
the $v_1$ is a rational function of
the variables $(b_1, \dots, b_L, v) \in (\R_{>0})^{L+1}$ 
and the parameters $s,l \in \R_{>0}$, because
it is given by a composition of rational maps.
Therefore,
by regarding the $b_i$'s also as parameters, we obtain an
algebraic equation $v = v_1$ for the unknown $v$ that assures the
periodic boundary condition.
Then we have:
%
%
\begin{proposition}\label{lem:1}
For any $s,l \in \R_{>0}$ and $(b_1, \dots, b_L) \in (\R_{>0})^{L}$,
there is a unique positive real solution $v \in \R_{>0}$ to the equation $v = v_1$. 
\end{proposition}
\proof
One observes that the $a'$ in \eqref{eq:jul28_1} 
is determined by the relation
\begin{equation*}
\begin{pmatrix}
b & l \\
1 & s/b
\end{pmatrix}
\begin{pmatrix}
a \\
1
\end{pmatrix}
= \left(a+\frac{s}{b} \right)
\begin{pmatrix}
a' \\
1
\end{pmatrix}.
\end{equation*}
Hence for \eqref{eq:jul28_2} we have
\begin{equation}\label{eq:nov19_1}
\begin{pmatrix}
L_{11} & L_{12} \\
L_{21} & L_{22}
\end{pmatrix}
\begin{pmatrix}
v \\
1
\end{pmatrix}
= \left(v+\frac{s}{b_L} \right) \left(v_L+\frac{s}{b_{L-1}} \right)
\cdots \left(v_2+\frac{s}{b_1} \right) 
\begin{pmatrix}
v_1 \\
1
\end{pmatrix},
\end{equation}
where the \textit{monodromy matrix} is given by
\begin{equation}\label{eq:sep16_1}
\begin{pmatrix}
L_{11} & L_{12} \\
L_{21} & L_{22}
\end{pmatrix}
:=
\begin{pmatrix}
b_1 & l \\
1 & s/b_1
\end{pmatrix}
\cdots
\begin{pmatrix}
b_L & l \\
1 & s/b_L
\end{pmatrix}.
\end{equation}
\textit{(Uniqueness.)}
Suppose there exist positive real solutions to the equation $v=v_1$.
From \eqref{eq:nov19_1} we see that for any such solution $v$ it is necessary for
$(v,1)^t$ to be a positive eigenvector
of the monodromy matrix \eqref{eq:sep16_1}.
Then by taking a ratio of the the components of
the vectors in both sides of
the equation \eqref{eq:nov19_1},
we obtain
\begin{equation}\label{eq:aug19_1}
v = \frac{L_{11} v + L_{12}}{L_{21} v + L_{22}}.
\end{equation}
%
Since this equation
has a unique positive real solution
\begin{equation}\label{eq:jul31_1}
v = \frac{L_{11} - L_{22} + \sqrt{(L_{11}- L_{22} )^2+4 L_{12} L_{21} }}{ 2 L_{21} },
\end{equation}
such a solution is unique.
\par\noindent
\textit{(Existence.)}
Equation \eqref{eq:nov19_1} is valid for any $v \in \R_{>0}$
so in particular for the $v$ in \eqref{eq:jul31_1}.
On the other hand, for this $v$ we also have
\begin{equation}\label{eq:nov25_1}
\begin{pmatrix}
L_{11} & L_{12} \\
L_{21} & L_{22}
\end{pmatrix}
\begin{pmatrix}
v \\
1
\end{pmatrix}
= (L_{21}v + L_{22})
\begin{pmatrix}
v \\
1
\end{pmatrix}.
\end{equation}
By equating the right hand side of the equation \eqref{eq:nov19_1} 
with that of \eqref{eq:nov25_1},
we see that this $v$ is indeed a solution to the algebraic equation $v=v_1$.
\qed

\par
With this $v \in \R_{>0}$ in \eqref{eq:jul31_1},
define $T_l: (\R_{>0})^L \rightarrow  (\R_{>0})^L$ to be a map given by
\begin{equation}\label{eq:aug4_1}
T_l (b_1, \dots, b_L) = (b'_1, \dots, b'_L),
\end{equation}
where the right hand side is determined by the relation
$\mathcal{R} (b_1, \dots, b_L, v) = (v, b'_1, \dots, b'_L)$.
We call this map a \textit{time evolution}, and $v$ a \textit{carrier} for the state $(b_1, \dots, b_L)$ 
associated with $T_l$. 
For any fixed $s \in \R_{>0}$,
now we obtained a one parameter family of
discrete time dynamical systems on the space $(\R_{>0})^L$
with the time evolutions $T_l  \, (l \in \R_{>0})$.
We would like to call such a system a closed geometric crystal chain.

As a discrete dynamical system,
the closed geometric crystal chain
has such properties that any
homogeneous state $(b_1, \dots, b_L) = (\alpha, \dots, \alpha)$
is a fixed point,
and for the case of even $L$
any alternating state $(b_1, \dots, b_L) = (\alpha, \beta, \dots, \alpha, \beta)$
is a periodic point with period $2$.
The latter one is related to a special modulo $2$ conservation law in
Remark \ref{rem:dec28_1}.
Also note that the time evolution $T_{s}$ produces
a cyclic shift by one spacial unit.

Here we present an example of the time evolutions of the closed geometric crystal chain.
Figure \ref{fig:1}  shows three results of
repeated applications of $T_l$s to an initial state of the system.
%
The patterns are showing that there are many `solitons' 
traveling with various velocities.
Also, one can observe that many
collisions of the solitons and many phase shifts induced by the collisions
are occurring in those patterns.
Such phenomena are typical to any dynamical system
with both non-linearity and integrability, including the periodic box-ball system.
We will give an additional observation on this example
at the end of section \ref{sec:2}.

\subsection{Properties of the dynamical system}\label{sec:2_2}
\subsubsection{Commutativity of the time evolutions.}\label{sec:2_2_1}
Since the geometric $R$-matrices satisfy the Yang-Baxter relation,
the following standard argument assures the commutativity of
the time evolutions.
Let $(b''_1, \dots, b''_L)= T_{l_2} \circ T_{l_1}(b_1, \dots, b_L)$
with the associated carriers $v$ (resp.~$\tilde{v}$) for the time evolutions
$T_{l_1}$ (resp.~$T_{l_2}$).
Then we have the relation
\begin{equation}\label{eq:aug24_1}
(R_2^{(s,l_2)} \circ \dots \circ R_{L+1}^{(s,l_2)}) \circ
(R_1^{(s,l_1)} \circ \dots \circ R_{L}^{(s,l_1)})(b_1, \dots, b_L,v,\tilde{v})=
(v,\tilde{v},b''_1, \dots, b''_L).
 \end{equation}
\clearpage
\vspace{2cm}
\par\noindent
\begin{figure}[htbp]
\centering
\includegraphics[height=8cm]{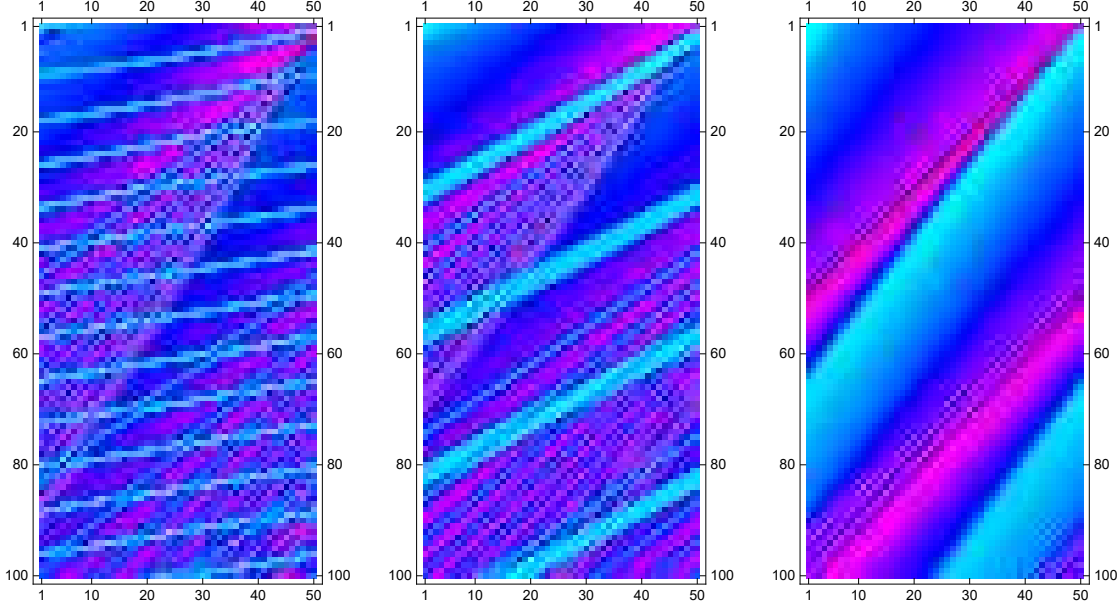}
\caption{Time evolutions of the closed geometric crystal chain for $n=2$ and $L=50$. 
The values of the parameters are $s=1$ (All), and
$l=0.00001$ (Left), $l=0.1$ (Middle), $l=2.0$ (Right).  
The initial state is given by $b_i = i/10 \quad (1 \leq i \leq 50)$
at the top row, and
the time flows from top to bottom.
Visualization is produced by the command MatrixPlot in Mathematica$^{\scriptstyle \mbox{\scriptsize \textregistered}}$.}
\label{fig:1}
\end{figure}
\par\noindent
By repeated use of the Yang-Baxter relation $R_{i+1}^{(s,l_2)} R_{i}^{(s,l_1)}  R_{i+1}^{(l_2,l_1)} = R_{i}^{(l_2,l_1)} R_{i+1}^{(s,l_1)} R_{i}^{(s,l_2)} $ and the involution
$R_{L+1}^{(l_2,l_1)} \circ R_{L+1}^{(l_1,l_2)} = {\rm Id}$, we obtain
\begin{align*}
&(R_2^{(s,l_2)} \circ \dots \circ R_{L+1}^{(s,l_2)}) \circ
(R_1^{(s,l_1)} \circ \dots \circ R_{L}^{(s,l_1)}) \\
&\quad = R_2^{(s,l_2)} \circ R_1^{(s,l_1)} \circ R_3^{(s,l_2)} \circ R_2^{(s,l_1)} \circ\dots \circ R_{L+1}^{(s,l_2)} \circ R_{L}^{(s,l_1)}\\
&\quad = (R_2^{(s,l_2)} \circ R_1^{(s,l_1)} \circ R_3^{(s,l_2)} \circ R_2^{(s,l_1)} \circ\dots \circ R_{L+1}^{(s,l_2)} \circ R_{L}^{(s,l_1)}) \circ (R_{L+1}^{(l_2,l_1)} \circ R_{L+1}^{(l_1,l_2)}) \\
&\quad =R_{1}^{(l_2,l_1)} \circ (R_2^{(s,l_1)} \circ R_1^{(s,l_2)} \circ R_3^{(s,l_1)} \circ R_2^{(s,l_2)} \circ \dots \circ R_{L+1}^{(s,l_1)} \circ R_{L}^{(s,l_2)}) \circ R_{L+1}^{(l_1,l_2)} \\
&\quad =R_{1}^{(l_2,l_1)} \circ (R_2^{(s,l_1)} \circ \dots \circ R_{L+1}^{(s,l_1)}) \circ
(R_1^{(s,l_2)} \circ \dots \circ R_{L}^{(s,l_2)})  \circ R_{L+1}^{(l_1,l_2)}.
\end{align*}
By substituting this into \eqref{eq:aug24_1} we obtain
\begin{equation}\label{eq:aug24_2}
(R_2^{(s,l_1)} \circ \dots \circ R_{L+1}^{(s,l_1)}) \circ
(R_1^{(s,l_2)} \circ \dots \circ R_{L}^{(s,l_2)})(b_1, \dots, b_L,u,\tilde{u})=
(u,\tilde{u},b''_1, \dots, b''_L),
 \end{equation}
where $(u,\tilde{u}) = R^{(l_1,l_2)} (v,\tilde{v})$.
Therefore we have
$T_{l_1} \circ T_{l_2} (b_1, \dots, b_L) = (b''_1, \dots, b''_L)$.
\subsubsection{Conservation laws.}\label{sec:2_2_2}
Here we show that the closed geometric crystal chains are
discrete integrable systems with $L/2$ (resp.~$(L+1)/2$)
conserved quantities for even (resp.~odd) $L$.

Given $a,b \in \R_{>0}$, the $a',b'$ in \eqref{eq:jul28_1} are 
determined by a pair of equations
\begin{equation}\label{eq:aug7_1}
ab = a'b', \qquad b+\frac{l}{a} = a'+\frac{s}{b'}. 
\end{equation}
It is equivalent to the following matrix equation
\begin{equation}
\begin{pmatrix}
b & \lambda \\
1 & s/b
\end{pmatrix}
\begin{pmatrix}
a & \lambda \\
1 & l/a
\end{pmatrix}
=
\begin{pmatrix}
a' & \lambda \\
1 & l/a'
\end{pmatrix}
\begin{pmatrix}
b' & \lambda \\
1 & s/b'
\end{pmatrix},
\end{equation}
where $\lambda$ is a parameter called the \textit{loop parameter} \cite{F19}.
Thus for the choice of $v$ in \eqref{eq:jul31_1},
the equation depicted by \eqref{eq:jul28_2} is written as
\begin{equation}\label{eq:jul31_2}
\begin{pmatrix}
b_1 & \lambda \\
1 & s/b_1
\end{pmatrix}
\cdots
\begin{pmatrix}
b_L & \lambda \\
1 & s/b_L
\end{pmatrix}
\begin{pmatrix}
v &  \lambda \\
1 &  l/v
\end{pmatrix}
= 
\begin{pmatrix}
v &  \lambda \\
1 &  l/v
\end{pmatrix}
\begin{pmatrix}
b'_1 & \lambda \\
1 & s/b'_1
\end{pmatrix}
\cdots
\begin{pmatrix}
b'_L & \lambda \\
1 & s/b'_L
\end{pmatrix}.
\end{equation}
Define $g(\alpha,\beta; \lambda)$ and 
$\mathcal{L}(| b \rangle; \lambda)$
for $| b \rangle := (b_1,\dots,b_L)$
to be $2 \times 2$ matrices given by
\begin{equation}\label{eq:dec28_2}
g(\alpha,\beta; \lambda) :=
\begin{pmatrix}
\alpha &  \lambda \\
1 &  \beta/\alpha
\end{pmatrix},
\quad
\mathcal{L}(| b \rangle; \lambda)=
\begin{pmatrix}
L_{11}(\lambda) & L_{12}(\lambda) \\
L_{21}(\lambda) & L_{22}(\lambda)
\end{pmatrix}
:=
g(b_1,s; \lambda) \cdots g(b_L,s; \lambda).
\end{equation}
Then equation \eqref{eq:jul31_2} is written as
\begin{equation}\label{eq:dec4_1}
\mathcal{L}(T_l | b \rangle; \lambda) = 
g(v,l; \lambda)^{-1} 
\mathcal{L}(| b \rangle; \lambda) g(v,l; \lambda),
\end{equation}
which can be viewed as a discrete time analogue of the
Lax equation \cite{Suris04}.
This implies that the characteristic polynomial
\begin{equation*}
\det (x \mathbb{I}_2 - \mathcal{L}(| b \rangle; \lambda))=
x^2 - (L_{11}(\lambda) + L_{22}(\lambda))x + \det \mathcal{L}(| b \rangle; \lambda)
\end{equation*}
is invariant under the time evolution $T_l$ for any $l \in \R_{>0}$.
Since $\det \mathcal{L}(| b \rangle; \lambda) = (s-\lambda)^L$ is trivially conserved,
all the non-trivial conserved quantities of this dynamical system are
contained in the trace $L_{11}(\lambda) + L_{22}(\lambda)$.

Let $b_j^{(1)}=b_j, b_j^{(2)}=s/b_j$ and for any $m \in \Z$ we 
extend its definition by $b_j^{(m)}=b_j^{(m-2)}$.
Based on \cite{ILP17, LP11}, define
the loop elementary symmetric functions $e^{(r)}_m (| b \rangle) \, (r=0,1)$ by
\begin{align}
e^{(1)}_m (| b \rangle)&= \sum_{1 \leq j_1 <j_2 < \dots < j_m \leq L}
b_{j_1}^{(2-j_1)} b_{j_2}^{(3-j_2)} \cdots b_{j_m}^{(1+m-j_m)}, \nonumber\\
e^{(0)}_m (| b \rangle)&= \sum_{1 \leq j_1 <j_2 < \dots < j_m \leq L}
b_{j_1}^{(1-j_1)} b_{j_2}^{(2-j_2)} \cdots b_{j_m}^{(m-j_m)},\label{eq:nov11_1}
\end{align}
and 
$e^{(r)}_0 (| b \rangle)=1,\, e^{(r)}_m (| b \rangle)=0 \, (m<0)$.
By Lemma 6.1 of \cite{ILP17}, we have
\begin{align}
L_{11}(\lambda) = \sum_{m \geq 0} e^{(1)}_{L-2m} (| b \rangle) \lambda^m,\quad
&L_{12}(\lambda) = \sum_{m > 0} e^{(1)}_{L+1-2m} (| b \rangle) \lambda^m,\nonumber\\
L_{21}(\lambda) = \sum_{m \geq 0} e^{(0)}_{L-1-2m} (| b \rangle) \lambda^m,\quad
&L_{22}(\lambda) = \sum_{m \geq 0} e^{(0)}_{L-2m} (| b \rangle) \lambda^m.
\label{eq:aug19_2}
\end{align}
Since the parameter $\lambda$ can take arbitrary values, 
every coefficient
\begin{equation}\label{eq:nov11_2}
I_{L-2m}:= e^{(1)}_{L-2m} (| b \rangle) + e^{(0)}_{L-2m} (| b \rangle),
\end{equation}
in the polynomial $L_{11}(\lambda) + L_{22}(\lambda)$
is a conserved quantity.
To summarize, there are $(L+1)/2$ conserved quantities
$I_1, I_3,\dots, I_L$ for odd $L$,
and $L/2$ conserved quantities $I_2, I_4,\dots, I_L$ for even $L$,
besides the trivial one $I_0=2$.
\begin{example}
The traces $L_{11}(\lambda) + L_{22}(\lambda)$ for up to $L=4$ are
as follows:
\begin{align}
L=1: &\quad b_1 + \bar{b}_1,\\
L=2: &\quad 2\lambda + b_1b_2+\bar{b}_1\bar{b}_2,\\
L=3: &\quad \lambda (b_1+\bar{b}_1+b_2+\bar{b}_2+b_3+\bar{b}_3)+b_1b_2b_3+\bar{b}_1\bar{b}_2\bar{b}_3,\\
L=4: &\quad 2 \lambda^2+\lambda ( b_1b_2+b_1\bar{b}_3+b_1b_4+\bar{b}_2\bar{b}_3+
\bar{b}_2b_4+b_3b_4 \nonumber\\
&\quad +\bar{b}_1\bar{b}_2
+\bar{b}_1b_3
+\bar{b}_1\bar{b}_4
+b_2b_3
+b_2 \bar{b}_4+
\bar{b}_3 \bar{b}_4 )
+b_1 b_2 b_3 b_4+\bar{b}_1\bar{b}_2\bar{b}_3\bar{b}_4.
\end{align}
Here $\bar{b}_i$ denotes $s/b_i$.
\end{example}

By taking $\lambda = 0$ in \eqref{eq:jul31_2}, we
see that $I'_L := e^{(1)}_{L} (| b \rangle) = b_1\dots b_L$ is also a conserved quantity,
which can be used in place of $I_L$.
\begin{remark}\label{rem:dec28_1}
In the case of even $L$, the
quantity $I'_1:=e^{(1)}_{1} (| b \rangle) = b_1+\bar{b}_2 +\dots +b_{L-1}+\bar{b}_L$ is invariant under any two consecutive time 
evolutions
$T_{l_2} \circ T_{l_1} (b_1, \dots, b_L) = (b''_1, \dots, b''_L)$.
This claim is verified by considering the equation
$\mathcal{L}(| b \rangle; \lambda) g(v,l_1; \lambda) g(\tilde{v},l_2; \lambda) =
g(v,l_1; \lambda) g(\tilde{v},l_2; \lambda) \mathcal{L}(| b'' \rangle; \lambda)$,
divided by $\lambda^{L/2+1}$ and then by taking the limit
$\lambda \rightarrow \infty$.
Since $e^{(1)}_{1} (| b \rangle) e^{(0)}_{1} (| b \rangle) = Ls + I_2$
is a conserved quantity,
the quantity $\bar{I}'_1:=e^{(0)}_{1} (| b \rangle)=\bar{b}_1+b_2+\dots + \bar{b}_{L-1}+b_L$
has also this property.
\end{remark}

\subsubsection{Invertibility.}\label{sec:2_2_3}
The time evolution $T_l$ defined in \eqref{eq:aug4_1} is invertible.
Actually, given any $(b_1,\dots,b_L) \in (\R_{>0})^L$ one can obtain
$(T_l)^{-1} (b_1, \cdots, b_L) = (\tilde{b}_1, \cdots, \tilde{b}_L)$
in the following way.
In view of \eqref{eq:jul31_2} we begin with the matrix equation
\begin{equation}\label{eq:aug4_2}
\begin{pmatrix}
\tilde{b}_1 & \lambda \\
1 & s/\tilde{b}_1
\end{pmatrix}
\cdots
\begin{pmatrix}
\tilde{b}_L & \lambda \\
1 & s/\tilde{b}_L
\end{pmatrix}
\begin{pmatrix}
\tilde{v} &  \lambda \\
1 &  l/\tilde{v}
\end{pmatrix}
= 
\begin{pmatrix}
\tilde{v} &  \lambda \\
1 &  l/\tilde{v}
\end{pmatrix}
\begin{pmatrix}
L_{11}(\lambda) & L_{12}(\lambda) \\
L_{21}(\lambda) & L_{22}(\lambda)
\end{pmatrix},
\end{equation}
where $\tilde{v}$ and $\tilde{b}_i$'s are the unknowns.
By flipping the matrices with respect to their anti-diagonals,
we see that this equation is equivalent to
\begin{equation}\label{eq:aug4_3}
\begin{pmatrix}
l/\tilde{v} &  \lambda \\
1 &  \tilde{v}
\end{pmatrix}
\begin{pmatrix}
s/\tilde{b}_L & \lambda \\
1 & \tilde{b}_L
\end{pmatrix}
\cdots
\begin{pmatrix}
s/\tilde{b}_1 & \lambda \\
1 & \tilde{b}_1
\end{pmatrix}
= 
\begin{pmatrix}
L_{22}(\lambda) & L_{12}(\lambda) \\
L_{21}(\lambda) & L_{11}(\lambda)
\end{pmatrix}
\begin{pmatrix}
l/\tilde{v} &  \lambda \\
1 &  \tilde{v}
\end{pmatrix}.
\end{equation}
In the same way as in Proposition \ref{lem:1} to obtain \eqref{eq:jul31_1},
we see that the $\tilde{v}$ satisfying this matrix equation 
is determined by the following equation
\begin{equation*}
l/\tilde{v}  = \frac{L_{22} l/\tilde{v}  + L_{12}}{L_{21} l/\tilde{v}  + L_{11}},
\end{equation*}
where $L_{ij} = L_{ij}(l)$s are given by \eqref{eq:sep16_1}.
Its unique positive real solution is
\begin{equation}
\tilde{v}  = \frac{ 2 l L_{21} }{L_{22} - L_{11} + \sqrt{(L_{11}- L_{22} )^2+4 L_{12} L_{21} }}.
\end{equation}
With this choice of $\tilde{v}$ we can obtain the $(\tilde{b}_1, \dots, \tilde{b}_L)$
in \eqref{eq:aug4_2} by using the inverse map 
$\mathcal{R}^{-1} = R_L \circ \cdots \circ R_1$ as
$\mathcal{R}^{-1} (\tilde{v}, b_1, \dots, b_L) = (\tilde{b}_1, \dots, \tilde{b}_L, \tilde{v})$.
\subsection{Continuum limits and associated differential equations}\label{sec:2_3}
\subsubsection{A naive method.}\label{sec:2_3_1}
In order to observe a few of the properties of our new discrete dynamical system, 
we consider two different continuum limits of the system.
First we note that this system has an obvious scale invariance. 
That is, by
replacing $a,b,s,l$ in \eqref{eq:jul28_1} by $\mu a, \mu b, \mu^2 s, \mu^2 l$
with a parameter $\mu \in \R_{>0}$ results in the replacements of
$a', b'$ by $\mu a', \mu b'$.
So, we may set $s=1$, and
we also use the letter $\tau$ instead of $l$.
Under this setting, we rewrite the equations in \eqref{eq:aug7_1} as
\begin{equation}\label{eq:aug7_2}
ab = a'b', \qquad b+\frac{\tau}{a} = a'+\frac{1}{b'}. 
\end{equation}

The first method is a naive one
in which we respect neither the integrability nor the periodicity
of the system.
Let $u(x,t), v(x,t)$ be a pair of variables depending on time $t$ and position $x$,
and $\delta > 0$ a small variable.
We set
\begin{align*}
a &= v(x + c_0 \delta,t), \quad b= u(x, t -\delta), \\
a' &= v(x - c_0 \delta,t), \quad b'= u(x, t +\delta) ,
\end{align*}
and require that both equations in \eqref{eq:aug7_2} are satisfied up to order $1$
of the variable $\delta$.
This requirement is satisfied if the variables $u(x,t), v(x,t)$ satisfy the 
following differential equation
\begin{equation}
\frac{\partial_t u(x,t)}{u(x,t)} = c_0 \frac{\partial_x v(x,t)}{v(x,t)},
\end{equation}
and if they are expressed by a new variable $U(x,t)$ as
\begin{align}
&v(x,t) = \sinh \left[ U(x,t) \right]+\sqrt{\tau  + \sinh^2 \left[ U(x,t) \right]},\\
&u(x,t) = \exp \left[ U(x,t) \right] .
\end{align}
Putting them together we obtain the following partial differential equation
\begin{equation}
\frac{1}{c_0} \frac{\partial U(x,t)}{\partial t}=
\frac{\cosh \left[ U(x,t) \right]}{\sqrt{\tau  + \sinh^2 \left[ U(x,t) \right]}}
 \frac{\partial U(x,t)}{\partial x}.
\end{equation}
This is a sort of nonlinear advection equation.
When $\tau = 1$ it reduces the linear differential equation $\partial_t U = c_0 \partial_x U$
for left moving waves of a common velocity $c_0$,
in agreement with the cyclic shift behavior of the original discrete dynamical system
for the case of time evolution $T_{s}$.
\subsubsection{Another method to respect the integrability.}\label{sec:2_3_2}
Here we assume that $L$ is an odd integer. 
According to diagram \eqref{eq:jul28_2} we write the evolution equation as
\begin{equation}\label{eq:sep1_1}
v_{i+1} b_i = v_{i} b'_i, \qquad b_i+\frac{\tau}{v_{i+1}} = v_i+\frac{1}{b'_i}. 
\end{equation}
Let $a_i(t), u_i(t)$ be a pair of variables depending on time $t$ and position $i \in \Z/L\Z$,
$C$ a constant,
and $\delta > 0$ a small variable.
We set
\begin{align*}
v_{i+1} &= \frac{1}{\delta} + a_{i+1}(t), \quad b_i= u_i(t), \\
v_i &= \frac{1}{\delta} + a_{i}(t), \qquad b'_i= u_i(t +\delta),\\
\tau &= \frac{1}{\delta^2} + \frac{C}{\delta},
\end{align*}
and write \eqref{eq:sep1_1} as the following discrete time analogue of the
Lax triads \cite{Suris04}
\begin{equation}
g(b_i',1;\lambda) = (\delta \cdot g(v_i, \tau; \lambda))^{-1} g(b_i,1;\lambda) 
(\delta \cdot g(v_{i+1}, \tau; \lambda)),
\end{equation}
where $g(\bullet, \bullet;\lambda)$ 
is the $2 \times 2$ matrix defined in \eqref{eq:dec28_2}.
Since 
\begin{math}
\delta \cdot g(v_i, \tau; \lambda) = \mathbb{I}_2 + \delta \cdot h(a_i(t),C;\lambda)
+ \mathcal{O} (\delta^2)
\end{math}
where
\begin{equation}
h(\alpha,\beta;\lambda) =
\begin{pmatrix}
\alpha & \lambda \\
1 & \beta - \alpha
\end{pmatrix},
\end{equation}
one can derive the following (continuous time) Lax triads  \cite{Suris04}
\begin{equation}\label{eq:dec4_2}
\frac{{\rm d}}{{\rm d} t} g(u_i(t),1;\lambda) = g(u_i(t),1;\lambda)h(a_{i+1}(t),C;\lambda)
-h(a_i(t),C;\lambda)g(u_i(t),1;\lambda).
\end{equation}
This implies that
\begin{align*}
a_{i+1}(t)+a_i(t) &= u_i(t) - \frac{1}{u_i(t)} + C, \\
\frac{{\rm d}}{{\rm d} t} u_i(t) &= u_i(t) (  a_{i+1}(t) - a_i(t)  ).
\end{align*}
Since $L$ is odd, we can solve the first 
couple of equations as
\begin{equation*}
a_i(t) = \frac12 \sum_{j=0}^{L-1} (-1)^j \left( u_{i+j}(t) - \frac{1}{u_{i+j}(t)} \right) +\frac{C}{2}.
\end{equation*}
By substituting this expression in the second equation, we have
\begin{equation*}
\frac{{\rm d}}{{\rm d} t} u_i(t) = u_i(t) \sum_{j=1}^{L-1} (-1)^{j-1} \left( u_{i+j}(t) - \frac{1}{u_{i+j}(t)} \right).
\end{equation*}
This may be viewed as a variation of the Lotka-Volterra equation.
If we set $u_i(t) = \exp[ U_i(t) ]$, the equation is written as
\begin{equation}
\frac{{\rm d}}{{\rm d} t} U_i(t) = 2 \sum_{j=1}^{L-1} (-1)^{j-1} \sinh U_{i+j}(t).
\end{equation}
This system
has obvious conserved quantities $\sum_{j=1}^{L}U_i(t)$ and
$\sum_{j=1}^{L} \cosh U_i(t)$.

A continuous limit of the discrete Lax equation 
\eqref{eq:dec4_1} is obtained by a standard method.
Let $\mathcal{L}(t) = g(u_1(t),1;\lambda) \cdots g(u_L(t),1;\lambda)$
and $\mathcal{B}(t) = h(a_1(t),C;\lambda)$.
Then we obtain
\begin{equation}
\frac{{\rm d}}{{\rm d} t} \mathcal{L}(t) = [\mathcal{L}(t), \mathcal{B}(t)],
\end{equation}
from the Lax triads equation \eqref{eq:dec4_2}.

\subsection{Tropicalization and piecewise linear formulas}\label{sec:2_4}
\subsubsection{An equation for periodic boundary conditions.}\label{sec:2_4_1}
Tropicalization is a procedure for turning subtraction-free rational maps 
$(\R_{>0})^{d_1} \rightarrow (\R_{>0})^{d_2}$ into piecewise-linear maps
$\R^{d_1} \rightarrow \R^{d_2}$ by replacing the operations $+,\cdot,\div$ with
the operations min, $+,-$, and ignoring constants.
In fact, there are some variations of the notion of tropicalization.
We adopt one of them which was described in \cite{IKT12}.
With an infinitesimal parameter $\varepsilon >0$,
define $\mathrm{Log}_\ve : \R_{>0} \to \R$  to be a map
given by
\begin{align}
  \label{i:loge-map}
  \mathrm{Log}_\ve : a \mapsto - \ve \log a.
\end{align}
For $a, b > 0$ define $\tp{a},\tp{b} \in \R$ by $a = \e^{-\frac{\tp{a}}{\ve}}$ and 
$b = \e^{-\frac{\tp{b}}{\ve}}$. 
Then we have  
$$
  \mathrm{Log}_\ve (a + b) = 
  -\ve \log (\e^{-\frac{\tp{a}}{\ve}} + \e^{-\frac{\tp{b}}{\ve}}),
  \quad
  \mathrm{Log}_\ve (a \times b) = \tp{a} + \tp{b}.
$$
In the limit $\ve \to 0$, $\mathrm{Log}_\ve (a + b)$ becomes $\min(\tp{a},\tp{b})$.
In this manner, the algebra $(\R_{>0},+,\times)$ reduces to 
the so called min-plus algebra,
and the procedure $\lim_{\ve \to 0} \mathrm{Log}_\ve$ 
with the transformation as $a = \e^{-\frac{\tp{a}}{\ve}}$
turns out to be the above mentioned tropicalization.

For example, the map of geometric $R$-matrix \eqref{eq:jul28_1}
is tropicalized to the following piecewise linear map
\begin{align}
A' &= A + \min (B,\tp{l}-A)-\min (A,\tp{s}-B),\nonumber\\
B'&= B+\min (A,\tp{s}-B)-\min (B,\tp{l}-A),
\end{align}
where $A = \tp{a}, B=\tp{b}$ and so on.
When we let the values of the variables be restricted to $\Z_{\geq 0}$, 
this reduces to the simplest case of
the combinatorial $R$-matrix in Kashiwara's crystal.
It is described by one-row tableaux of two kinds of letters (1 and 2)
\begin{equation}
\overbrace{1\dots1}^{B} \overbrace{2\dots2}^{\tp{s}-B} \otimes
\overbrace{1\dots1}^{A} \overbrace{2\dots2}^{\tp{l}-A} \mapsto
\overbrace{1\dots1}^{A'} \overbrace{2\dots2}^{\tp{l}-A'}  \otimes
\overbrace{1\dots1}^{B'} \overbrace{2\dots2}^{\tp{s}-B'} .
\end{equation}
The generalized periodic box-ball system \cite{KS08, T09}
may be regarded as a tropicalization of the closed geometric
crystal chain in section \ref{sec:2_1}.
Compared with the diagram \eqref{eq:jul28_2},
the corresponding situation may be depicted by
\begin{equation}\label{eq:nov12_3}
\batten{V_1}{B_1}{B_1'}{V_2}\!\!\!
\batten{}{B_2}{B_2'}{V_3}\!\!\!
\batten{}{}{}{\cdots\cdots}
\quad
\batten{}{}{}{V_{L-1}}\,\,
\batten{}{B_{L-1}}{B_{L-1}'}{V_{L}}\!\!\!
\batten{}{B_L}{B_L'}{V,}
\end{equation}
where $V=\tp{v}$, $B_i = \tp{b_i}$ and so on.
However, the assertion of Lemma \ref{lem:1} which 
tells the existence of a unique solution to the equation
$V_1=V$,
does not persist in the tropicalization.
This is due to the fact that the expression for $v$ in \eqref{eq:jul31_1} 
is not subtraction-free rational. 
Therefore, to study the periodic boundary condition for the 
generalized periodic box-ball system,
we have to consider the tropicalization of the equation \eqref{eq:aug19_1}
itself.
It reads as
\begin{equation}\label{eq:aug20_1}
V = \min (\Lambda_{11} + V, \Lambda_{12}) - \min( \Lambda_{21} + V, \Lambda_{22}).
\end{equation}
Here 
\begin{equation}\label{eq:nov13_1}
\Lambda_{ij} = \tp{L_{ij}} = \tp{L_{ij}(l)},
\end{equation}
that are tropicalizations of the matrix elements of
the monodromy matrix \eqref{eq:sep16_1}.
They can be explicitly written down by using
the loop elementary symmetric functions 
\eqref{eq:nov11_1} and \eqref{eq:aug19_2}.
\begin{example}\label{ex:nov12_1}
The $\Lambda_{11}, \Lambda_{21}$ for up to $L=4$ are
as follows:
\begin{align*}
L=1: &\quad \Lambda_{11}=B_1, \Lambda_{21}=0,\\
L=2: &\quad \Lambda_{11}=\min (\tp{l},B_1+B_2), \Lambda_{21}=\min (\bar{B}_1, B_2),\\
L=3: &\quad \Lambda_{11}=\min (\tp{l}+\min(B_1,\bar{B}_2,B_3), B_1+B_2+B_3), \nonumber\\
&\quad \Lambda_{21}=\min (\tp{l}, \bar{B}_1+\bar{B}_2, \bar{B}_1+B_3, B_2+B_3),\\
L=4: &\quad \Lambda_{11}=\min \left( 2\tp{l}, \tp{l}+\min(B_1+B_2,B_1+\bar{B}_3,B_1+B_4,\right. \nonumber\\
&\quad \quad \quad \quad \quad \quad  
\left. \bar{B}_2+\bar{B}_3, \bar{B}_2+B_4, B_3+B_4 \right),
B_1+B_2+B_3+B_4), \nonumber\\
&\quad \Lambda_{21}=\min (\tp{l}+\min(\bar{B}_1,B_2,\bar{B}_3,B_4),\nonumber\\
&\quad \quad \quad \quad \quad \quad  \bar{B}_1+\bar{B}_2+\bar{B}_3,
\bar{B}_1+\bar{B}_2+B_4,\bar{B}_1+B_3+B_4,B_2+B_3+B_4).
\end{align*}
Here $\bar{B}_i$ denotes $\tp{s}- B_i$.
The other matrix elements are given by $\Lambda_{22}=\Lambda_{11}(B_i \leftrightarrow \bar{B}_i), \Lambda_{12}=\tp{l}+\Lambda_{21}(B_i \leftrightarrow \bar{B}_i)$.
\end{example}
The following result is easily obtained by a simple case-by-case check. 
\begin{proposition}\label{pr:nov12_2}
The solution to the equation \eqref{eq:aug20_1} is given by:
\begin{enumerate}
\item If $\frac{\Lambda_{12}+\Lambda_{21}}{2} \leq \min (\Lambda_{11},\Lambda_{22})$,
then $V=\frac{\Lambda_{12}-\Lambda_{21}}{2}$.
\item If $\Lambda_{11} < \min (\Lambda_{22}, \frac{\Lambda_{12}+\Lambda_{21}}{2} )$,
then $V=\Lambda_{11}-\Lambda_{21}$.
\item If $\Lambda_{22} < \min (\Lambda_{11}, \frac{\Lambda_{12}+\Lambda_{21}}{2} )$,
then $V=\Lambda_{12}-\Lambda_{22}$.
\item If $\Lambda_{11}=\Lambda_{22} < \frac{\Lambda_{12}+\Lambda_{21}}{2}$,
then any $V$ such that $\Lambda_{11}-\Lambda_{21} \leq V \leq \Lambda_{12}-\Lambda_{11}$.
\end{enumerate}
\end{proposition}
\proof
For simplicity, let $A=\Lambda_{11}, B=\Lambda_{12}, C=\Lambda_{21}$ and
$D= \Lambda_{22}$.
\par\noindent
Case (i): Suppose $C+V>D$. Then we have $V>D-C \geq \frac{B+C}{2}-C=
 \frac{B-C}{2}$,
hence $A+V > \frac{B+C}{2}+\frac{B-C}{2}=B$.
So by \eqref{eq:aug20_1} we get $V=B-D$, but this leads to
$D < C+V = C+B-D \leq 2D -D =D$, a contradiction.
Thus $C+V \leq D$.
Suppose $A+V<B$, which implies $V=A-C$ by \eqref{eq:aug20_1}. 
But this leads to $B > A+V = 2A-C \geq B+C-C =B$, a contradiction.
Therefore $A+V \geq B$, hence by \eqref{eq:aug20_1} 
we get $V=\frac{B-C}{2}$.
\par\noindent
Case (ii): Suppose $A+V>B$.
Then if $C+V \geq D$ we get $V=B-D$ by \eqref{eq:aug20_1},
which leads to $B < A+V = A+B-D < B$, a contradiction.
Otherwise we have $C+V < D$ and then $V=\frac{B-C}{2}$ by \eqref{eq:aug20_1},
which leads to $B < A+V < \frac{B+C}{2}+\frac{B-C}{2}=B$, a contradiction.
Thus $A+V \leq B$.
Then if $C+V > D$ we have $A=D$ by \eqref{eq:aug20_1}
that contradicts to the assumption.
Therefore $C+V \leq D$, hence by \eqref{eq:aug20_1} 
we get $V=A-C$.
\par\noindent
Case (iii): A proof can be obtained from the previous case by exchanging
$A$ with $D$, $B$ with $C$, and $V$ with $-V$.
\par\noindent
Case (iv): Suppose $A+V>B$.
Then if $C+V > D$ we get $V=B-D$ by \eqref{eq:aug20_1},
which leads to $B < A+V = A+B-D = B$, a contradiction.
Otherwise we have $C+V \leq D$ and then $V=\frac{B-C}{2}$ by \eqref{eq:aug20_1},
which leads to $D \geq C+V = \frac{B+C}{2}>D$, a contradiction.
Thus $A+V \leq B$.
Then if $C+V < D$ we have $V=A-C$ by \eqref{eq:aug20_1},
which leads to $D > C+V = A =D$, a contradiction.
Therefore $C+V \geq D$, hence by \eqref{eq:aug20_1} 
we get $A=D$ which does not contradict to the assumption.
Thus any $V$ satisfying the condition $A-C \leq V \leq B-A$ 
solves the equation \eqref{eq:aug20_1}.
\qed

Here we show two examples to examine this result.
\begin{example}\label{ex:nov13_2}
Set $L=2, \tp{s}=2$, and $B_1 = B_2 =1$.
This is the state $12 \otimes 12$ in tableau notation.
By Example \ref{ex:nov12_1} one has
$\Lambda_{11} = \Lambda_{22} = \min (\tp{l},2)$,
and $\frac{\Lambda_{12}+\Lambda_{21}}{2}= 1+\tp{l}/2$.
Hence it falls into either case (i) when $\tp{l}=2$
or into case (iv) otherwise.
In any case the solution is given by $\min (\tp{l}-1,1) \leq V \leq \max (\tp{l}-1,1)$.
In particular, when $\tp{l}=1$ we have $V=0$ and $V=1$ as possible
solutions of integer values.
Then the diagram \eqref{eq:nov12_3} reads as
\begin{equation*}
\batten{1}{1}{0}{0}\!\!\!\!\!\!
\batten{}{1}{2}{1}
\qquad
\begin{picture}(40,40)(-20,-20)
	\put(0,-1){\makebox(0,0)[b]{and}}
\end{picture}
\qquad
\batten{0}{1}{2}{1}\!\!\!\!\!\!
\batten{}{1}{0}{0},
\end{equation*}
respectively.
The periodic boundary condition $V_1=V$ is indeed satisfied, but
the output states
($22 \otimes 11$ and $11 \otimes 22$ in tableau notation) are different. 
So, we can not define a unique time evolution compatible with such
periodic boundary conditions.
\end{example}

\begin{example}\label{ex:nov13_3}
Set $L=3, \tp{s}=2$, and $B_1 = B_2 =B_3=1$.
This is the state $12 \otimes 12 \otimes 12$  in tableau notation.
By Example \ref{ex:nov12_1} one has
$\Lambda_{11} = \Lambda_{22} = 1+ \min (\tp{l},2)$,
and $\frac{\Lambda_{12}+\Lambda_{21}}{2}= \min (\tp{l},2)+\tp{l}/2$.
Hence it falls into either case (i) when $\tp{l}=1,2$
or into case (iv) otherwise.
In the former case the solution is given by 
$V =\tp{l}/2$, and
in the latter case it is given by 
$1 \leq V \leq \tp{l}-1$.
In particular, when $\tp{l}=1$ we have $V=1/2$ as the solution, but
it is not an integer:
\begin{equation*}
\batten{\frac12}{1}{1}{\frac12}\!\!\!\!\!\!
\batten{}{1}{1}{\frac12}\!\!\!\!\!\!
\batten{}{1}{1}{\frac12}.
\end{equation*}
In fact, the only possible diagram \eqref{eq:nov12_3} 
for integer value $V$ is given by
\begin{equation*}
\batten{1}{1}{0}{0}\!\!\!\!\!\!
\batten{}{1}{2}{1}\!\!\!\!\!\!
\batten{}{1}{0}{0}
\qquad
\begin{picture}(40,40)(-20,-20)
	\put(0,-1){\makebox(0,0)[b]{or}}
\end{picture}
\qquad
\batten{0}{1}{2}{1}\!\!\!\!\!\!
\batten{}{1}{0}{0}\!\!\!\!\!\!
\batten{}{1}{2}{1},
\end{equation*}
so neither satisfies the periodic boundary condition.
\end{example}

In the generalized periodic box-ball system, there are states
that do not admit time evolutions by carriers with
specific capacities \cite{KS08}.
The above two examples show how
such `non-evolvable' states actually appear.

\subsubsection{Conservation laws and the energy of paths.}\label{sec:2_4_2}
Although
the closed geometric crystal chain itself cannot be tropicalized in the sense that
the expression for the carrier $v$ in \eqref{eq:jul31_1} 
is not subtraction-free rational,
its conserved quantities are given by polynomials with non-negative integer 
coefficients and hence can be tropicalized.
It is fairly reasonable to expect that
the tropicalizations of these conserved quantities are
the conserved quantities of the generalized periodic box-ball systems in 
\cite{KS08}.

The tropicalization of the
loop elementary symmetric functions \eqref{eq:nov11_1} are given by
\begin{align*}
\tp{e^{(1)}_m (| b \rangle)}&= \min_{1 \leq j_1 <j_2 < \dots < j_m \leq L}
\left( B_{j_1}^{(2-j_1)} +B_{j_2}^{(3-j_2)} +\cdots +B_{j_m}^{(1+m-j_m)} \right), \\
\tp{e^{(0)}_m (| b \rangle)}&= \min_{1 \leq j_1 <j_2 < \dots < j_m \leq L}
\left( B_{j_1}^{(1-j_1)} +B_{j_2}^{(2-j_2)} +\cdots +B_{j_m}^{(m-j_m)} \right),
\end{align*}
where $B_{j}^{(r)}$ denotes $\tp{b_{j}^{(r)}}$.
Note that $B_{j}^{(r)} + B_{j}^{(r-1)} = \tp{s}$, and $r$ is interpreted in modulo $2$.
From the arguments in section \ref{sec:2_2_2} to deduce \eqref{eq:nov11_2},
it is reasonable to consider that the piecewise-linear functions
\begin{equation}
\tp{I_{L-2m}} = \min \left( \tp{e^{(1)}_{L-2m} (| b \rangle)}, 
\tp{e^{(0)}_{L-2m} (| b \rangle)} \right),
\end{equation}
with $m \in \{ 0,1,\ldots,\lfloor L/2 \rfloor \}$
provide a collection of conserved quantities of the
generalized periodic box-ball system for the following initial state or `path'
\begin{equation}\label{eq:oct23_1}
p = \overbrace{1\dots1}^{B_{1}^{(1)}} \overbrace{2\dots2}^{B_{1}^{(2)}} \otimes
\overbrace{1\dots1}^{B_{2}^{(1)}} \overbrace{2\dots2}^{B_{2}^{(2)}} \otimes
\cdots \otimes
\overbrace{1\dots1}^{B_{L}^{(1)}} \overbrace{2\dots2}^{B_{L}^{(2)}}.
\end{equation}
Based on the notion of an isospectral evolution, we also want to 
find an explicit piecewise-linear expression for the tropicalization of
the eigenvalues of the monodromy matrix \eqref{eq:sep16_1}
that is identical with the matrix $\mathcal{L}(| b \rangle; l)$.
To this end, we first consider the trace of this matrix
\begin{equation}\label{eq:oct23_2}
\tp{L_{11}(l) + L_{22}(l)}
=\min_{m \in \{ 0,1,\ldots,\lfloor L/2 \rfloor \}}
\left( m \tp{l}+\tp{I_{L-2m}} \right) = 
\min(\Lambda_{11}, \Lambda_{22}).
\end{equation}
Here the last expression is due to \eqref{eq:nov13_1}.
%
Then the piecewise-linear functions \eqref{eq:oct23_2} with $\tp{l}=1,2,\dots$
also provide a collection of conserved quantities 
for the initial state \eqref{eq:oct23_1}.
We note that there is an inequality
\begin{equation}\label{eq:oct28_2}
\tp{L_{11}(l) + L_{22}(l)} \leq \tp{I_L} = \min\left( 
\sum_{i=1}^L B_{i}^{(1)}, \sum_{i=1}^L B_{i}^{(2)}
\right) \leq
\frac{L}{2} \tp{s}.
\end{equation}
Now we consider the roots of the quadratic equation
\begin{equation*}
\det (x \mathbb{I}_2 - \mathcal{L}(| b \rangle; l))=
x^2 - (L_{11}(l) + L_{22}(l))x +(s-l)^L =0.
\end{equation*}
One of the roots is the Perron-Frobenius eigenvalue $E_l (>0)$ of 
the positive matrix $\mathcal{L}(| b \rangle; l)$.
As we will see in section \ref{sec:3_2_5}, we can regard $\tp{E_l}$ as
the \textit{energy of path} \cite{KS08} for the initial state \eqref{eq:oct23_1} of the generalized periodic box-ball system.
\begin{proposition}
We have the following formula
\begin{equation}\label{eq:oct28_1}
\tp{E_l} =  \min \left( \tp{L_{11}(l) + L_{22}(l)} , L \frac{\tp{l}}{2} \right).
\end{equation}
\end{proposition}

\proof
First we consider the case where $L$ is even or $s \geq l$.
Denote the other root by $F_l$.
Then we have $F_l \geq 0$. 
If $F_l >0$ then one can tropicalize $F_l$ as well as $E_l$.
Since $E_l$ is the Perron-Frobenius eigenvalue
we have $E_l > F_l$, hence $\tp{E_l} < \tp{F_l}$.
Therefore we can obtain a piecewise linear formula
\begin{equation}
\tp{E_l} = \min \left( \tp{E_l}, \tp{F_l} \right) = \tp{L_{11}(l) + L_{22}(l)}.
\end{equation}
Obviously, this result is also valid for the case of $F_l = 0$.
Then if $L$ is even, this is equivalent to \eqref{eq:oct28_1}
which can be verified by \eqref{eq:oct23_2} and the fact $\tp{I_0}=0$.
Otherwise we have $\tp{s} \leq \tp{l}$ and $L$ is odd,
hence by the inequality \eqref{eq:oct28_2} we obtain the same result. 

Next we consider the case where $L$ is odd and $s < l$.
Apply the map $\mathrm{Log}_\ve$ on both sides of the equation
\begin{equation*}
E_l^2 = (L_{11}(l) + L_{22}(l))E_l +(l-s)^L,
\end{equation*}
and take the limit $\ve \to 0$.
Then by noting that
\begin{equation*}
\mathrm{Log}_\ve (l-s)^L =
L \left( \tp{l} - \ve \log \left( 1-\e^\frac{\tp{l} -\tp{s}}{\ve} \right) \right),
\end{equation*}
and $\tp{l} -\tp{s}<0$, we can derive the following piecewise linear equation
\begin{equation}
2 \tp{E_l} = \min \left( \tp{L_{11}(l) + L_{22}(l)} + \tp{E_l}, L \tp{l} \right).
\end{equation}
It is easy to see that this equation has a unique solution \eqref{eq:oct28_1}.
\qed

Since $\tp{E_l}$ is a function of $\tp{l}$, we let
$\mathcal{E}_{\tp{l}}$ denote $\tp{E_l}$.
The notion of the \textit{number of the solitions of length $j$}
was first introduced in \cite{FOY00} for non-periodic box-ball systems,
and then also for periodic systems \cite{KTT}.
In our notation for the tropicalized energy of path,
it is given by
\begin{equation}
m_j = -\mathcal{E}_{j-1} + 2 \mathcal{E}_{j} + \mathcal{E}_{j+1}.
\end{equation}
In the case of the generalized periodic box-ball system \cite{KS08},
this collection of numbers $\{ m_j\}_{j=1,2,\dots}$ were defined only
for evolvable paths.
In contrast, by using the formula \eqref{eq:oct28_1} we can formally define this
quantity even for non-evolvable paths.

For any path of the form $p$ in \eqref{eq:oct23_1}, define its weight to be
$\wt (p) = \sum_{i=1}^L (B_i^{(1)}-B_i^{(2)})$.
In the following examples, we restrict ourselves to
consider the paths with non-negative weights.
\begin{example}
Set $L=2, \tp{s}=2$.
Then we have $\mathcal{E}_{j} = \min (B_1+B_2, \bar{B}_1+\bar{B}_2, j)$.
There are one path with no solitons $11 \otimes 11$,
two paths with one soliton of length one, $11 \otimes 12, 12 \otimes 11$,
and three paths with one soliton of length two, 
$11 \otimes 22, 22 \otimes 11, 12 \otimes 12$.
The last one is the non-evolvable path in Example \ref{ex:nov13_2}.
\end{example}

\begin{example}
Set $L=3, \tp{s}=2$.
Then we have $\mathcal{E}_{j} = 
\min (B_1+B_2+B_3, \bar{B}_1+\bar{B}_2+\bar{B}_3, 
j+\min(B_1,\bar{B}_1,B_2, \bar{B}_2,B_3,\bar{B}_3), 3j/2)$.
There are one path with no solitons $11 \otimes 11 \otimes 11$,
three paths with one soliton of length one, $11 \otimes 11 \otimes 12, 
11 \otimes 12 \otimes 11, 12 \otimes 11 \otimes 11$,
six paths with one soliton of length two
\begin{align*}
&11 \otimes 11 \otimes 22, 11 \otimes 22 \otimes 11, 22 \otimes 11 \otimes 11,\\
&11 \otimes 12 \otimes 12, 12 \otimes 12 \otimes 11, 12 \otimes 11 \otimes 12,
\end{align*}
six paths with one soliton of length three
\begin{align*}
&11 \otimes 12 \otimes 22, 12 \otimes 22 \otimes 11, 22 \otimes 11 \otimes 12,\\
&11 \otimes 22 \otimes 12, 22 \otimes 12 \otimes 11, 12 \otimes 11 \otimes 22,
\end{align*}
and one path of `three halves' solitons of length two $12 \otimes 12 \otimes 12$. 
The last one, which has fractional number of solitons, 
is the non-evolvable path in Example \ref{ex:nov13_3}.
\end{example}

In the box-ball systems, $\tp{l}$ is interpreted as the capacity of
a carrier that carries the balls.
When the time evolution of the box-ball system is given by
a carrier with the capacity $\tp{l}$, it is known that
a soliton of length $j$ has a constant velocity $\min (j, \tp{l})$ when 
the soliton is
sufficiently separated from the other solitons. 
So if we let $\tp{l}$ be more and more larger,
differences of the speeds of the solitons due to their lengths
become more and more larger.
Note that larger $\tp{l}$ implies smaller $l$.
Actually,
in Figure \ref{fig:1} in section \ref{sec:2_1} we observe that 
differences of the speeds of the `solitons'
in the case for $l=0.00001$ are larger than those
in the case for $l=0.1$. 

\section{The case of general $\bm{n}$}\label{sec:3}
\subsection{Totally one-row tableaux case}\label{sec:3_1}
\subsubsection{Definition of the dynamical system.}\label{sec:3_1_1}
Based on the notions in \cite{F19, F18}, we 
introduce the \textit{positive real} rational $1$-rectangle by $\mathbb{Y}_1 = 
(\R_{>0})^{n-1} \times \R_{>0}$.
Let $(\mathbf{x}, s)$ denote
an element of $\mathbb{Y}_1$
with $\mathbf{x} = (x^{(1)},\dots,x^{(n-1)})$, and set $x^{(n)}:=s/(x^{(1)} \cdots x^{(n-1)})$.
Furthermore, we define $x^{(i)}$ for arbitrary $i \in \Z$ to be a variable determined
from $\mathbf{x}$ by the relation  $x^{(i)} = x^{(i+n)}$.

Let $E_{i,j}$ be an $n \times n$ matrix which has $1$ in the $(i, j)$ position and
$0$ elsewhere.
Given a fixed loop parameter $\lambda$, we define
\begin{equation}
\Lambda_j(\alpha_1,\dots,\alpha_n) = \sum_{i=1}^{n-j} \alpha_i E_{i+j,i} + 
\lambda \sum_{i=n-j+1}^n \alpha_i E_{i+j-n,i},
\end{equation}
for $j \in \{ 0,1,\dots,n-1\}$.
For any $(\mathbf{x}, s) \in \mathbb{Y}_1$, let $g(\mathbf{x}, s; \lambda)$ denote
the associated unipotent crystal matrix defined to be
\begin{equation}\label{eq:sep29_1}
g(\mathbf{x}, s; \lambda) = \Lambda_0(x^{(1)},\dots,x^{(n)}) +
\Lambda_1(1,\dots,1).
\end{equation}
From its $(n-1)$-th minor determinants we define another $n \times n$
matrix $g^*(\mathbf{x}, s; \lambda)$ as
\begin{equation}
g^*(\mathbf{x}, s; \lambda) = \sum_{j=0}^{n-1} \Lambda_j 
\left(
\prod_{i=1}^{n-j-1} x^{(i)}, \prod_{i=1}^{n-j-1} x^{(i-1)}, \dots,\prod_{i=1}^{n-j-1} x^{(i-n+1)} 
\right).
\end{equation}
\begin{example}\label{ex:nov20_2}
In the case of $n=4$ these matrices look like
\begin{align*}
g(\mathbf{x}, s; \lambda) &=
\begin{pmatrix}
x^{(1)} &0 & 0& \lambda \\
1 & x^{(2)} &0 &0 \\
 0& 1 & x^{(3)} & 0\\
0& 0& 1 & x^{(4)} 
\end{pmatrix},\\
g^*(\mathbf{x}, s; \lambda) &= 
\begin{pmatrix}
x^{(1)}  x^{(2)} x^{(3)}& \lambda &  \lambda x^{(3)}&  \lambda x^{(2)} x^{(3)}  \\
x^{(1)}  x^{(2)}  &x^{(4)}  x^{(1)}  x^{(2)} &  \lambda&   \lambda x^{(2)}\\
x^{(1)}  & x^{(4)}  x^{(1)}   & x^{(3)} x^{(4)}  x^{(1)}  & \lambda \\
1 &x^{(4)}  &  x^{(3)} x^{(4)} &x^{(2)} x^{(3)} x^{(4)} 
\end{pmatrix}.
\end{align*}
These matrices are shifted and folded
versions of the ``whirl'' and the ``curl'' in \cite{LP12}.\end{example}

In order to explain the definition of the matrix $g^*(\mathbf{x}, s; \lambda)$,
here we introduce a useful notion. 
For any $n \times n$ matrix $A$ and $1 \leq r \leq n$, let $\mathsf{C}_r (A)$ be
the \textit{$r$-th contravariant alternating tensor representation} of $A$, which is 
an ${n \choose r} \times {n \choose r}$
matrix that consists of all the order $r$ minor determinants of $A$ 
(See, for example \cite{Satake73}).
That is, we define
\begin{equation}\label{eq:sep2_2}
\mathsf{C}_r (A) = \{ \Delta_{I,J} (A) \}_{I,J \in {[n] \choose r}},
\end{equation}
where the indices are assumed to be in lexicographic order if they are
regarded as words, e.g.~$i_1 i_2 \ldots i_r$ for $I = \{ i_1< i_2<\dots< i_{r} \}$.
Then we have 
\begin{equation}\label{eq:nov13_5}
\mathsf{C}_{n-1} (g(\mathbf{b}, s; (-1)^n\lambda)) =  g^*(\mathbf{b}, s; \lambda).
\end{equation}

Now we consider the following matrix equation
\begin{equation}\label{eq:sep2_1}
g(\mathbf{b}, s; \lambda) g(\mathbf{a}, l; \lambda) =
g(\mathbf{a}', l; \lambda) g(\mathbf{b}', s; \lambda).
\end{equation}
For any $s, l \in \R_{>0}$ and
$(\mathbf{a},\mathbf{b}) \in (\R_{>0})^{2n-2} $, 
there is a unique solution 
$(\mathbf{a}',\mathbf{b}') \in (\R_{>0})^{2n-2} $ to 
this matrix equation
(\cite{F18}, and see also Remark \ref{rem:nov19_3} for the case of $s=l$).
Let
$R^{(s,l)}: (\R_{>0})^{2n-2}  \rightarrow  (\R_{>0})^{2n-2} $ be a rational map
given by $R^{(s,l)}:(\mathbf{b},\mathbf{a}) \mapsto (\mathbf{a}',\mathbf{b}')$.
This is the geometric $R$-matrix in the present case, and
if we write
$\mathbf{a} = (a^{(1)},\dots,a^{(n-1)}), \mathbf{b} = (b^{(1)},\dots,b^{(n-1)})$
and so on, an explicit expression for the solution is given by  \cite{LP12, Y01}
\begin{equation}\label{eq:nov19_2}
a'^{(j)} = a^{(j)} \frac{\kappa_{j+1}}{\kappa_j}, \quad
b'^{(j)} = b^{(j)} \frac{\kappa_j}{\kappa_{j+1}}, 
\end{equation}
where
\begin{equation}
\kappa_j = \kappa_j(\mathbf{b},\mathbf{a})=
\sum_{r=0}^{n-1} a^{(j)} \cdots a^{(j+r-1)}  b^{(j+r+1)} \cdots b^{(j+n-1)}. 
\end{equation}
From the definition of the geometric $R$-matrix in \cite{F18},
we see that:
\begin{lemma}\label{lem:2}
The elements of $\mathbf{a}' \in (\R_{>0})^{n-1} $ are determined by the
following formula
\begin{equation}\label{eq:aug28_1}
\begin{pmatrix}
a'^{(1)} a'^{(2)}  \cdots \cdots a'^{(n-1)} \\
a'^{(1)} a'^{(2)}  \cdots a'^{(n-2)} \\
\dots \\
a'^{(1)}  a'^{(2)}  \\
a'^{(1)}   \\
1 
\end{pmatrix}
=
\frac{1}{\kappa_1(\mathbf{b},\mathbf{a})}
g^*(\mathbf{b}, s; l)
\begin{pmatrix}
a^{(1)} a^{(2)}  \cdots \cdots a^{(n-1)} \\
a^{(1)} a^{(2)}  \cdots a^{(n-2)} \\
\dots \\
a^{(1)}  a^{(2)}  \\
a^{(1)}   \\
1 
\end{pmatrix}.
\end{equation}
\end{lemma}
\proof
For any $n \times n$ matrix $A$, we denote by $\pi (A)$ the $n \times (n-1)$
matrix obtained from $A$ by dropping its last column.
Let $\overline{\Theta}_{n-1} (\mathbf{a})$ and $\overline{\Theta}_{n-1} (\mathbf{a}')$ be the $n-1$ dimensional subspaces
of $\C^n$ spanned by the columns of $\pi (g(\mathbf{a}, l; \bullet) )$
and $\pi (g(\mathbf{a}', l; \bullet) )$, respectively.
We regard them as elements of the Grassmannian ${\rm Gr}(n-1,n)$.
Then by the definition of the geometric $R$-matrix
(Definition 5.1 of \cite{F18}), they are related by
$\overline{\Theta}_{n-1} (\mathbf{a}') = g(\mathbf{b}, s; (-1)^n l) \cdot \overline{\Theta}_{n-1} (\mathbf{a})$,
where the meaning of
$\cdot$ in the right hand side was given in section \ref{sec:1_3}.
As a matrix representative of $\overline{\Theta}_{n-1} (\mathbf{a}')$, we introduce an $n \times (n-1)$
matrix $\tilde{M} (\mathbf{a}')$ given by
$\tilde{M} (\mathbf{a}') = g(\mathbf{b}, s; (-1)^n l) \pi (g(\mathbf{a}, l; \bullet) )$.
Then we have
\begin{equation}
\frac{P_{I}(\overline{\Theta}_{n-1} (\mathbf{a}'))}{P_{J}(\overline{\Theta}_{n-1} (\mathbf{a}'))}=
\frac{\Delta_{I, [n-1]}(\tilde{M} (\mathbf{a}'))}{\Delta_{J, [n-1]}
(\tilde{M} (\mathbf{a}'))},
\end{equation}
for any $(n-1)$-subsets $I, J$ of $[n]$.

It is easy to see that the elements of $\mathbf{a}'$ are given by ratios of the Pl$\ddot{\rm u}$cker  
coordinates of $\overline{\Theta}_{n-1} (\mathbf{a}')$.
More explicitly, 
their expressions are given by
 $a'^{(i)} = P_{[n] \setminus \{ i+1 \}}(\overline{\Theta}_{n-1} (\mathbf{a}'))/P_{[n] \setminus \{ i \}}(\overline{\Theta}_{n-1} (\mathbf{a}'))$.
Therefore the $i$-th element of the left hand side of equation \eqref{eq:aug28_1}
is given by
\begin{equation}
a'^{(1)} a'^{(2)}  \cdots a'^{(n-i)}=
\frac{P_{[n] \setminus \{ n-i+1 \}}(\overline{\Theta}_{n-1} (\mathbf{a}'))}{P_{[n] \setminus \{ 1 \}}(\overline{\Theta}_{n-1} (\mathbf{a}'))}=
\frac{\Delta_{[n] \setminus \{ n-i+1 \}, [n-1]}(\tilde{M} (\mathbf{a}'))}{\Delta_{[n] \setminus \{ 1 \}, [n-1]}
(\tilde{M} (\mathbf{a}'))}.
\end{equation}
By the Cauchy-Binet formula, we can write its numerator as
\begin{equation}
\Delta_{[n] \setminus \{ n-i+1 \}, [n-1]}(\tilde{M} (\mathbf{a}')) =
\sum_{j=1}^n g^*(\mathbf{b}, s; l)_{i,j} g^*(\mathbf{a}, l; \bullet)_{j,1},
\end{equation}
where we used \eqref{eq:nov13_5}. 
On the other hand,
by the formulas for the geometric coenergy functions
(Definition 6.3 and Corollary 7.3 of \cite{F18}) we can write its denominator as
\begin{equation}
\Delta_{[n] \setminus \{ 1 \}, [n-1] }
(\tilde{M} (\mathbf{a}'))= 
\kappa_1(\mathbf{b},\mathbf{a}).
\end{equation}
The proof is completed.
\qed
\begin{remark}\label{rem:nov19_3}
When $s=l$, the matrix equation \eqref{eq:sep2_1} has
a trivial solution $\mathbf{a}' = \mathbf{b}, \mathbf{b}' = \mathbf{a}$.
In this case, the non-trivial solution \eqref{eq:nov19_2} reduces to
this trivial one.
This claim is verified by using the following formula
\begin{align*}
&g^*(\mathbf{b},s;s)\\
&\quad ={\rm diag} (\prod_{i=1}^{n-1} b^{(i)}, \dots, b^{(1)}b^{(2)},b^{(1)},1)
\begin{pmatrix}
1 & \dots & 1 \\
\vdots & & \vdots \\
1 & \dots & 1
\end{pmatrix}
{\rm diag} (1, b^{(n)},b^{(n-1)}b^{(n)},\dots,\prod_{i=2}^{n} b^{(i)}),
\end{align*} 
and Lemma \ref{lem:2}.
\end{remark}

As in the case of $n=2$,
we define $R_i^{(s,l)}$ and $\mathcal{R}^{(s,l)} = R_1^{(s,l)} \circ \cdots \circ R_L^{(s,l)}$
which are now maps from $ (\R_{>0})^{(L+1)(n-1)} $ to itself.
Given an arbitrary $(\mathbf{b}_1, \dots, \mathbf{b}_L, \mathbf{v}) \in  (\R_{>0})^{(L+1)(n-1)} $,
let $\mathcal{R}^{(s,l)} (\mathbf{b}_1, \dots, \mathbf{b}_L, \mathbf{v}) = 
(\mathbf{v}_1, \mathbf{b}'_1, \dots, \mathbf{b}'_L)$.
It is depicted by
\begin{equation}\label{eq:aug28_2}
\batten{\mathbf{v}_1}{\mathbf{b}_1}{\mathbf{b}_1'}{\mathbf{v}_2}\!\!\!
\batten{}{\mathbf{b}_2}{\mathbf{b}_2'}{\mathbf{v}_3}\!\!\!
\batten{}{}{}{\cdots\cdots}
\quad
\batten{}{}{}{\mathbf{v}_{L-1}}\,\,
\batten{}{\mathbf{b}_{L-1}}{\mathbf{b}_{L-1}'}{\mathbf{v}_{L}}\!\!\!
\batten{}{\mathbf{b}_L}{\mathbf{b}_L'}{\mathbf{v} \in (\R_{>0})^{(n-1)}.}
\end{equation}
Once again,
we would like to construct a
discrete time dynamical system on the space $ (\R_{>0})^{L(n-1)}$
using a map that sends $(\mathbf{b}_1, \dots, \mathbf{b}_L)$ to $(\mathbf{b}'_1, \dots, \mathbf{b}'_L)$
as a unit step of its time evolution.
We see that
the 
$\mathbf{v}_1$ at the left end is a rational function of
the variables $(\mathbf{b}_1, \dots, \mathbf{b}_L, \mathbf{v}) 
\in  (\R_{>0})^{(L+1)(n-1)}$ 
and the parameters $s,l \in \R_{>0}$,
because it is given by a composition of rational maps.
Again, by regarding the $\mathbf{b}_i$'s also as parameters, we obtain an 
algebraic equation $\mathbf{v} = \mathbf{v}_1$ for the unknown $\mathbf{v}$
that assures the system of having a
periodic boundary condition.
Then we have:

\begin{proposition}\label{lem:3}
For any $s,l \in \R_{>0}$ and $(\mathbf{b}_1, \dots, \mathbf{b}_L) 
\in  (\R_{>0})^{L(n-1)} $,
there is a unique positive real solution $\mathbf{v} \in(\R_{>0})^{n-1}$ to the 
equation $\mathbf{v} = \mathbf{v}_1$.
\end{proposition}
\proof
Let $| \mathbf{b} \rangle = (\mathbf{b}_1, \dots, \mathbf{b}_L)$
and $\mathsf{M}_l^{(1)}(| \mathbf{b} \rangle) = g^*(\mathbf{b}_{1}, s; l) \cdots g^*(\mathbf{b}_{L}, s; l)$ which we call a monodromy matrix.
By repeated use of Lemma \ref{lem:2} we have
\begin{align}
&\mathsf{M}_l^{(1)}(| \mathbf{b} \rangle)
\begin{pmatrix}
v^{(1)} v^{(2)}  \cdots \cdots v^{(n-1)} \\
v^{(1)} v^{(2)}  \cdots v^{(n-2)} \\
\dots \\
v^{(1)}  v^{(2)}  \\
v^{(1)}   \\
1 
\end{pmatrix}=\kappa_1(\mathbf{b}_{L},\mathbf{v})
\cdots
\kappa_1(\mathbf{b}_{1},\mathbf{v}_{2})
\begin{pmatrix}
v_1^{(1)} v_1^{(2)}  \cdots \cdots v_1^{(n-1)} \\
v_1^{(1)} v_1^{(2)}  \cdots v_1^{(n-2)} \\
\dots \\
v_1^{(1)}  v_1^{(2)}  \\
v_1^{(1)}   \\
1 
\end{pmatrix}.\nonumber\\
&
\label{eq:nov19_4}
\end{align}
By the Perron-Frobenius theorem, there exists a positive 
eivenvector $\vec{\xi} = (\xi_1,\xi_2,\dots,\xi_n)^t$ of
the positive matrix $\mathsf{M}_l^{(1)}(| \mathbf{b} \rangle)$,
and it is unique up to a scalar multiple.
By using this $\vec{\xi}$,
define $\mathbf{v}= (v^{(1)},\dots,v^{(n-1)})  \in(\R_{>0})^{n-1}$ to be a vector
given by
\begin{equation}\label{eq:nov19_5}
v^{(1)} = \xi_{n-1}/\xi_{n}, v^{(2)} = \xi_{n-2}/\xi_{n-1}, \dots, 
v^{(n-1)} = \xi_{1}/\xi_{2}.
\end{equation}
Equation \eqref{eq:nov19_4} is valid for any $\mathbf{v} \in(\R_{>0})^{n-1}$
so in particular for this $\mathbf{v}$ given by \eqref{eq:nov19_5}.
On the other hand, by definition this unique $\mathbf{v}$ also satisfies
\begin{equation}\label{eq:nov25_2}
\mathsf{M}_l^{(1)}(| \mathbf{b} \rangle)
\begin{pmatrix}
v^{(1)} v^{(2)}  \cdots \cdots v^{(n-1)} \\
v^{(1)} v^{(2)}  \cdots v^{(n-2)} \\
\dots \\
v^{(1)}  v^{(2)}  \\
v^{(1)}   \\
1 
\end{pmatrix}=E^{(1)}_l
\begin{pmatrix}
v^{(1)} v^{(2)}  \cdots \cdots v^{(n-1)} \\
v^{(1)} v^{(2)}  \cdots v^{(n-2)} \\
\dots \\
v^{(1)}  v^{(2)}  \\
v^{(1)}   \\
1 
\end{pmatrix},
\end{equation}
where $E^{(1)}_l$ is the dominant (or Perron-Frobenius) eigenvalue
of the monodromy matrix.
By equating the right hand side of the equation \eqref{eq:nov19_4} with 
that of \eqref{eq:nov25_2},
we see that this $\mathbf{v}$ is a solution to the algebraic equation 
$\mathbf{v} = \mathbf{v}_1$, and that 
there is no other solution because $\vec{\xi}$ is unique.
\qed

With this unique solution $\mathbf{v}$ 
in Proposition \ref{lem:3} we
define $T^{(1)}_l:  (\R_{>0})^{L(n-1)}  \rightarrow  (\R_{>0})^{L(n-1)}$
to be a map given by
\begin{equation}\label{eq:aug30_1}
T^{(1)}_l (\mathbf{b}_1, \dots, \mathbf{b}_L) = (\mathbf{b}'_1, \dots, \mathbf{b}'_L),
\end{equation}
where the right hand side is determined by the relation
$\mathcal{R}^{(s,l)} (\mathbf{b}_1, \dots, \mathbf{b}_L, \mathbf{v}) = 
(\mathbf{v}, \mathbf{b}'_1, \dots, \mathbf{b}'_L)$.
We call this map a \textit{time evolution}, and $\mathbf{v}$ a \textit{carrier} for the state $(\mathbf{b}_1, \dots, \mathbf{b}_L)$ 
associated with $T^{(1)}_l$. 
This time evolution defines a family of discrete time dynamical systems
on the space $ (\R_{>0})^{L(n-1)}$.
As in the $n=2$ case, we call such a system a closed geometric crystal chain.

We note that, any homogeneous state is a fixed point of this dynamical system.
To verify this claim, consider the result of
Proposition \ref{lem:3} for the one site $L=1$ case.
Then by \eqref{eq:nov19_2}, we have $\mathbf{b}'_1 = \mathbf{b}_1$.
Let $\mathbf{v}$ be the carrier for the state $(\mathbf{b}_1)$.
Then we have $\mathcal{R}^{(s,l)} (\mathbf{b}_1, \dots, \mathbf{b}_1, \mathbf{v}) = 
(\mathbf{v}, \mathbf{b}_1, \dots, \mathbf{b}_1)$.
Therefore, this $\mathbf{v}$ is also the unique 
carrier for the state $(\mathbf{b}_1, \dots, \mathbf{b}_1)$ and we have
$T^{(1)}_l (\mathbf{b}_1, \dots, \mathbf{b}_1) = (\mathbf{b}_1, \dots, \mathbf{b}_1)$
for any $l \in \R_{>0}$.
Also note that the time evolution $T^{(1)}_{s}$ produces
a cyclic shift by one spacial unit.
This is a consequence of the claim in Remark \ref{rem:nov19_3}.

\subsubsection{Conservation laws.}\label{sec:3_1_2}
By combining the claims in Remark \ref{rem:nov19_3}
and Proposition \ref{lem:3}
we have the following:
\begin{theorem}\label{th:main}
For any $s,l \in \R_{>0}$ and $(\mathbf{b}_1, \dots, \mathbf{b}_L) 
\in  (\R_{>0})^{L(n-1)} $,
there is a unique positive real solution 
$(\mathbf{v}, \mathbf{b}'_1, \dots, \mathbf{b}'_L) \in(\R_{>0})^{(L+1)(n-1)}$ to the 
following matrix equation
\begin{equation}\label{eq:nov20_1}
g(\mathbf{b}_1, s; \lambda) \cdots g(\mathbf{b}_L, s; \lambda) g(\mathbf{v}, l; \lambda) =
g(\mathbf{v}, l; \lambda) g(\mathbf{b}'_1, s; \lambda) \cdots g(\mathbf{b}'_L, s; \lambda).
\end{equation}
\end{theorem}
We introduce an $n \times n$ matrix
$\mathcal{L}(| \mathbf{b} \rangle ; \lambda) $
for $| \mathbf{b} \rangle = (\mathbf{b}_1, \dots, \mathbf{b}_L)$
as follows
\begin{equation}\label{eq:sep29_2}
\mathcal{L}(| \mathbf{b} \rangle ; \lambda) =
g(\mathbf{b}_{1}, s; \lambda)
\cdots
g(\mathbf{b}_{L}, s; \lambda).
\end{equation}
We call this matrix a Lax matrix.
Due to the matrix equation \eqref{eq:nov20_1},
the time evolution \eqref{eq:aug30_1} is described by a
discrete time analogue of the Lax equation
\begin{equation}
\mathcal{L}(T^{(1)}_l | \mathbf{b} \rangle ; \lambda)=
g(\mathbf{v},l; \lambda)^{-1}
\mathcal{L}(| \mathbf{b} \rangle ; \lambda) g(\mathbf{v},l; \lambda).
\end{equation}

In order to study the characteristic polynomial of the Lax matrix,
here we present a well-known result related to the contravariant 
alternating tensor representation \eqref{eq:sep2_2}.
That is, the characteristic polynomial of any $n \times n$ matrix $A$ is given by
\begin{equation*}
\det (x \mathbb{I}_n - A)=
x^n +  \sum_{k=1}^{n-1} (-1)^{n-k} {\rm Tr}\mathsf{C}_{n-k}( A) 
x^{k}
+ (-1)^n\det A.
\end{equation*}
(This formula is derived by using calculus of a determinant. See, for example 
\cite{Satake73}.
An alternative derivation for the case of $A \in {\rm M}_n(\C)$ will be
given as a consequence of Corollary \ref{cor:oct12_2}.)
The Lax representation implies that the characteristic polynomial 
(and hence every coefficient therein) of
the matrix $\mathcal{L}(| \mathbf{b} \rangle ; \lambda)$
is invariant under the time evolution $T^{(1)}_l$ for any $l \in \R_{>0}$.
Therefore,
all the non-trivial conserved quantities of a closed geometric crystal chain are
contained in the traces ${\rm Tr}\mathsf{C}_{n-k}( \mathcal{L}(| \mathbf{b} \rangle ; \lambda)) = 
\sum_{I \in {[n] \choose n-k}} \Delta_{I,I} (\mathcal{L}(| \mathbf{b} \rangle ; \lambda)) $
for $1 \leq k \leq n-1$,
besides $\det \mathcal{L}(| \mathbf{b} \rangle ; \lambda) = (s+(-1)^{n-1}\lambda)^L$
which is trivially conserved.
More precisely, since each
${\rm Tr}\mathsf{C}_{n-k}( \mathcal{L}(| \mathbf{b} \rangle ; \lambda))$
is a polynomial of the loop parameter $\lambda$,
its all coefficients are separately conserved.
%

As in the case of $n=2$, there is an explicit expression for the matrix elements
in the $(i, j)$ position of the Lax matrix (Lemma 6.1 of \cite{ILP17}) given by
\begin{equation*}
\mathcal{L}(| \mathbf{b} \rangle ; \lambda)_{ij}
=\sum_{m \geq 0} e^{(i)}_{j-i+L-mn} (| \mathbf{b} \rangle) \lambda^m ,
\end{equation*}
where the loop elementary symmetric functions $e^{(r)}_m (| \mathbf{b} \rangle)$ are defined as
\begin{equation*}
e^{(r)}_m (| \mathbf{b} \rangle)= \sum_{1 \leq j_1 <j_2 < \dots < j_m \leq L}
b_{j_1}^{(r+1-j_1)} b_{j_2}^{(r+2-j_2)} \cdots b_{j_m}^{(r+m-j_m)},
\end{equation*}
and 
$e^{(r)}_0 (| \mathbf{b} \rangle)=1,\, e^{(r)}_m (| \mathbf{b} \rangle)=0 \, (m<0 \quad
\mbox{or} \quad s>L)$.
Therefore, an expression for the conserved quantities
of the closed geometric crystal chains is given by using
this explicit formula for the matrix elements to calculate
the minor determinants $\Delta_{I,I}(\mathcal{L}(| \mathbf{b} \rangle ; \lambda))$.
We will give a discussion on another expression for the conserved quantities
in section \ref{sec:4}.

\subsection{The case of rectangular tableaux for the carriers}\label{sec:3_2}
\subsubsection{Definition of the geometric $R$-matrices.}\label{sec:3_2_1}
\begin{figure}[htbp]
\centering
\includegraphics[height=8cm]{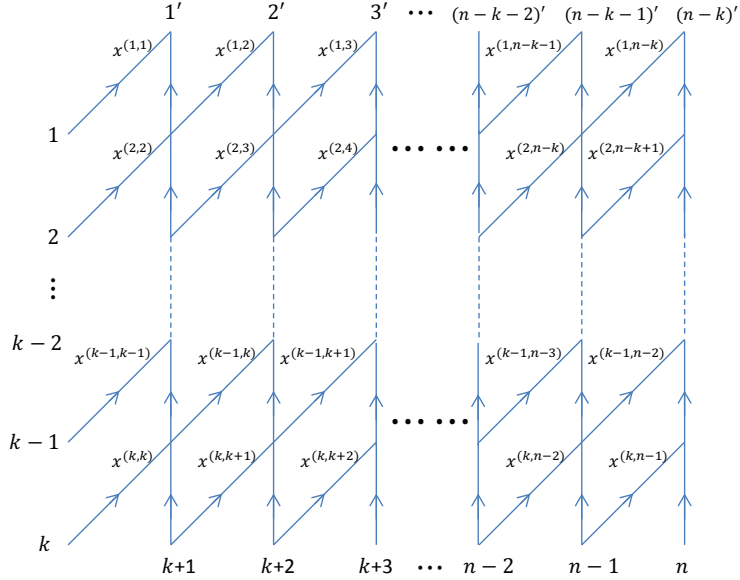}
\caption{The planar network $N(\mathbf{x})$. 
Vertical edges have weight 1.}
\label{fig:2}
\end{figure}

As in the one-row tableaux case, we
introduce the positive real rational $k$-rectangle by $\mathbb{Y}_k = 
\overline{\mathbb{Y}}_k \times \R_{>0}$, where $\overline{\mathbb{Y}}_k:=(\R_{>0})^{k(n-k)}$.
Define
\begin{equation}
R_k = \{ (i,j) \vert 1 \leq i \leq k,  i \leq j \leq i+n-k-1 \},
\end{equation}
to be an index set.
Let $(\mathbf{x}, l)$ denote an element of $\mathbb{Y}_k$
with $\mathbf{x} = (x^{(i,j)})_{(i,j) \in R_k}$, and set 
$x^{(i,i+n-k)}:=l/\prod_{j=i}^{i+n-k-1} x^{(i,j)}$ for $1 \leq i \leq k$.
In its associated rectangular tableau with $k$-rows,
$\tp{x^{(i,j)}}$ denotes the number of $j$'s in the $i$th row
and $\tp{l}$ denotes the width of the tableau.
By Definition 4.7 of \cite{F19}, we associate for such $\mathbf{x}$ a planar network $N(\mathbf{x})$ as in Figure \ref{fig:2}, where
diagonal edges are weighted by $x^{(i,j)}$ and vertical ones are by $1$.
%
It has $n$ sources labeled $1,\dots,n$ and $n-k$ sinks labeled $1',\dots,(n-k)'$.
Define the weight of a path to be the product of the weights of the edges in the
path.
Let $M(\mathbf{x})$ be an $n \times (n-k)$ matrix 
associated to the network $N(\mathbf{x})$, such that
whose $(i,j)$-entry is the sum of the weights of all paths from source $i$ to
sink $j'$.

\begin{example}\label{ex:dec22_1}
In the case of $n=4$ 
and $\mathbf{x}, \mathbf{y}, \mathbf{z}$ for $\overline{\mathbb{Y}}_{k=1,2,3}$
these matrices look like
\begin{equation*}
M(\mathbf{x}) =
\begin{pmatrix}
x^{(1,1)} & 0& 0 \\
1 & x^{(1,2)} &0 \\
 0& 1 & x^{(1,3)} \\
0&0 & 1 
\end{pmatrix},\quad
M(\mathbf{y}) =
\begin{pmatrix}
y^{(1,1)} & 0 \\
y^{(2,2)} & y^{(1,2)}y^{(2,2)}\\
1 & y^{(1,2)} + y^{(2,3)}  \\
0 & 1
\end{pmatrix},
\end{equation*}
and
\begin{equation*}
M(\mathbf{z}) =
\begin{pmatrix}
z^{(1,1)}  \\
z^{(2,2)}\\
z^{(3,3)} \\
1 
\end{pmatrix}.
\end{equation*}
These matrices are given by the planar networks
in Figure \ref{fig:2.55}.
\clearpage
\begin{figure}[htbp]
\centering
\includegraphics[height=5cm]{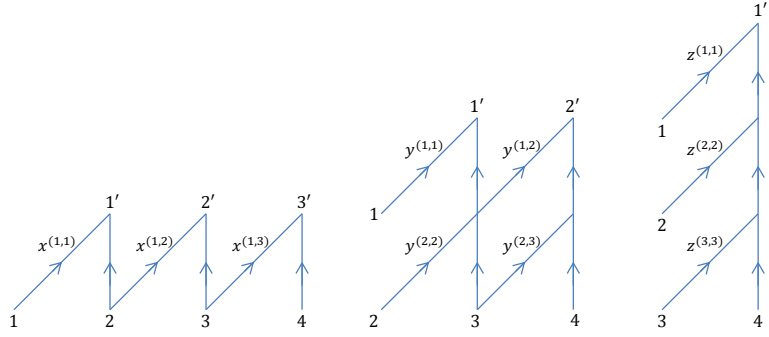}
\caption{The planar networks $N(\mathbf{x}), N(\mathbf{y}), N(\mathbf{z})$ 
for the matrices $M(\mathbf{x}), M(\mathbf{y}), M(\mathbf{z})$.}
\label{fig:2.55}
\end{figure}
\end{example}

Let $\overline{\Theta}_{n-k} (\mathbf{x})$ be the $n-k$ dimensional subspace
of $\C^n$ spanned by the columns of $M (\mathbf{x})$, which is
regarded as a point lies in the Grassmannian variety ${\rm Gr}(n-k,n)$.
This notation implies that there is a map $\overline{\Theta}_{n-k}: \overline{\mathbb{Y}}_k
\rightarrow {\rm Gr}(n-k,n)$, that is basically identical to the one
which is called the \textit{Gelfand-Tsetlin parametrization} 
of ${\rm Gr}(n-k,n) \times \C^{\times}$ in \cite{F19, F18}.

It is known that the \textit{positivity} of the Pl$\ddot{\rm u}$cker  
coordinates (Corollary 4.15 of \cite{F19}) holds,
which tells that for all $I \in {[n] \choose n-k}$, $P_I(\overline{\Theta}_{n-k} (\mathbf{x}))$
is a non-zero (homogeneous) polynomial in the quantities $x^{(i,j)}$
with non-negative integer coefficients.
For the currently using setup where we are dealing with
only positive real parameters and variables,
this implies that
for any $\mathbf{x} \in \overline{\mathbb{Y}}_k$
every $P_I(\overline{\Theta}_{n-k} (\mathbf{x}))$ takes its value in $\R_{>0}$.
In other words, $\overline{\Theta}_{n-k} (\mathbf{x})$ is a point
lies in the \textit{totally positive Grassmannian} 
${\rm Gr}(n-k,n)_{>0} \subset {\rm Gr}(n-k,n)$ \cite{L16}. 
%
Moreover, the map $\overline{\Theta}_{n-k}$
is a bijection between $\overline{\mathbb{Y}}_k$
and ${\rm Gr}(n-k,n)_{>0}$.
In fact, for $M \in {\rm Gr}(n-k,n)_{>0}$ its inverse image 
$\mathbf{x}= (x^{(i,j)})_{(i,j) \in R_k} = (\overline{\Theta}_{n-k})^{-1}(M)
\in \overline{\mathbb{Y}}_k$
is given by the following formula (Proposition 4.3 of \cite{F19})
\begin{equation}\label{eq:sep8_1}
x^{(i,i)} = \frac{P_{J_{i,i}}(M)}{P_{J_{i+1,i}}(M)}
\quad \mbox{or} \quad
x^{(i,j)} = \frac{P_{J_{i,j}}(M)P_{J_{i+1,j-1}}(M)}{P_{J_{i+1,j}}(M)P_{J_{i,j-1}}(M)},
\end{equation}
for $j \in [i+1, i+n-k-1]$.
Here we used the
notation for the \textit{basic subsets} 
$J_{i,j}:=[i,j] \cup [k+j-i+2,n] \in {[n] \choose n-k}$.

Let $g^{(n-k)}(\mathbf{x},l;\lambda)$ denote
an $n \times n$ matrix that is introduced in \cite{F19} to
define the unipotent crystal map.
The elements in the $(i,j)$ position of this matrix are defined by
\begin{equation*}
g^{(n-k)}(\mathbf{x},l;\lambda)_{ij}=c_{ij}
\frac{P_{[j-n+k+1,j-1] \cup \{i\}}(\overline{\Theta}_{n-k} (\mathbf{x}))}{P_{[j-n+k,j-1]}(\overline{\Theta}_{n-k} (\mathbf{x}))},
\quad
c_{ij} =
\begin{cases}
1 & \mbox{if } j \leq n-k, \\
l & \mbox{if } j > n-k \mbox{ and } i \geq j, \\
\lambda & \mbox{if } j > n-k \mbox{ and } i < j.
\end{cases}
\end{equation*}
\begin{example}\label{ex:dec22_2}
For the $\mathbf{x}, \mathbf{y}, \mathbf{z}$
in Example \ref{ex:dec22_1} for $n=4$ and
$\overline{\mathbb{Y}}_{k=1,2,3}$, we have
\begin{equation*}
g^{(3)}(\mathbf{x},l;\lambda) =
\begin{pmatrix}
\frac{P_{134}}{P_{234}} &0 & 0& \lambda \\
1 & \frac{P_{124}}{P_{134}} &0 &0\\
 0& 1 & \frac{P_{123}}{P_{124}} &0 \\
0& 0& 1 & l \frac{P_{234}}{P_{123}} 
\end{pmatrix}=
\begin{pmatrix}
x^{(1,1)} & 0& 0& \lambda \\
1 & x^{(1,2)} & 0&0 \\
0 & 1 & x^{(1,3)} & 0\\
0&0 & 1 & x^{(1,4)} 
\end{pmatrix},
\end{equation*}
\begin{equation*}
g^{(2)}(\mathbf{y},l;\lambda) =
\begin{pmatrix}
\frac{P_{14}}{P_{34}} &0 & \lambda & \lambda \frac{P_{13}}{P_{23}}\\
\frac{P_{24}}{P_{34}}  & \frac{P_{12}}{P_{14}} & 0 & \lambda \\
1 & \frac{P_{13}}{P_{14}}  & l \frac{P_{23}}{P_{12}}  & 0 \\
0&1 & l \frac{P_{24}}{P_{12}}  & l \frac{P_{34}}{P_{23}} 
\end{pmatrix}=
\begin{pmatrix}
y^{(1,1)} & 0 & \lambda & \lambda \frac{y^{(1,1)}(y^{(1,2)}+y^{(2,3)})}{y^{(2,2)}y^{(2,3)}} \\
y^{(2,2)} & y^{(1,2)}y^{(2,2)} & 0 & \lambda \\
1 & y^{(1,2)} + y^{(2,3)} & y^{(1,3)}y^{(2,3)} & 0 \\
0 & 1 & y^{(1,3)} & y^{(2,4)} 
\end{pmatrix},
\end{equation*}
and
\begin{equation*}
g^{(1)}(\mathbf{z},l;\lambda) =
\begin{pmatrix}
\frac{P_{1}}{P_{4}} &\lambda & \lambda \frac{P_{1}}{P_{2}}& \lambda \frac{P_{1}}{P_{3}}\\
\frac{P_{2}}{P_{4}} & l \frac{P_{2}}{P_{1}} & \lambda & \lambda \frac{P_{2}}{P_{3}}\\
\frac{P_{3}}{P_{4}} & l \frac{P_{3}}{P_{1}}  & l \frac{P_{3}}{P_{2}}  & \lambda \\
1& l \frac{P_{4}}{P_{1}}& l \frac{P_{4}}{P_{2}}  & l \frac{P_{4}}{P_{3}}
\end{pmatrix}=
\begin{pmatrix}
z^{(1,1)} & \lambda & \lambda \frac{z^{(1,1)}}{z^{(2,2)}}& \lambda \frac{z^{(1,1)} }{z^{(3,3)} } \\
z^{(2,2)} & z^{(1,2)}z^{(2,2)} & \lambda & \lambda \frac{z^{(2,2)}}{z^{(3,3)}} \\
z^{(3,3)} & z^{(1,2)}z^{(3,3)} & z^{(2,3)}z^{(3,3)} & \lambda \\
1 & z^{(1,2)} & z^{(2,3)} & z^{(3,4)} 
\end{pmatrix}.
\end{equation*}
Note that the first $n-k$ columns of $g^{(n-k)}(\mathbf{x},l;\lambda)$
coincide with those of
the matrix $M (\mathbf{x})$.
\end{example}
Several
properties of the matrix $g^{(n-k)}(\mathbf{x},l;\lambda)$ 
presented in \cite{F18}
(in our notations) are listed as follows.
\begin{proposition}[\cite{F18}: Proposition 3.17 and Corollary 3.18]\label{pr:nov27_1}
Let $(\mathbf{x}, l) \in \mathbb{Y}_k, A = g^{(n-k)}(\mathbf{x},l;\lambda)$,
and $B \in {\rm M}_n (\C [\lambda, \lambda^{-1}])$.
\begin{enumerate}
\item The first $n-k$ columns of $A$ span the subspace $\overline{\Theta}_{n-k} (\mathbf{x})$.
\item The matrix $A \vert_{\lambda = (-1)^{n-k-1}l}$ has rank $n-k$.
\item The determinant of $A$ is $(l + (-1)^{n-k} \lambda)^k$.
\item The first $n-k$ columns of $A \cdot B \vert_{\lambda = (-1)^{n-k-1}l}$
are contained in the subspace $\overline{\Theta}_{n-k} (\mathbf{x})$.
\end{enumerate}
\end{proposition}
Now we consider the following matrix equation
\begin{equation}\label{eq:sep10_3}
g^{(n-1)}(\mathbf{b}, s; \lambda) g^{(n-k)}(\mathbf{a}, l; \lambda) =
g^{(n-k)}(\mathbf{a}', l; \lambda) g^{(n-1)}(\mathbf{b}', s; \lambda).
\end{equation}
\begin{proposition}\label{pr:nov27_4}
For any $s, l \in \R_{>0}$ and
$(\mathbf{b},\mathbf{a}) \in \overline{\mathbb{Y}}_1 \times \overline{\mathbb{Y}}_k$, 
there is a unique solution 
$(\mathbf{a}',\mathbf{b}') \in \overline{\mathbb{Y}}_k \times \overline{\mathbb{Y}}_1$ to 
the matrix equation \eqref{eq:sep10_3}.
\end{proposition}
\proof
Besides differences of the notations,
an explicit expression for a solution to \eqref{eq:sep10_3}
is available
in section 7.1 of \cite{F18}, where the map 
that sends $(\mathbf{b},\mathbf{a})$ to $(\mathbf{a}',\mathbf{b}')$
is denoted by $\Theta R$.
It remains to prove the uniqueness of the solution.
Suppose there are two solutions $(\mathbf{a}',\mathbf{b}')$ and
 $(\mathbf{a}'',\mathbf{b}'')$, hence
\begin{equation}\label{eq:nov27_2}
g^{(n-k)}(\mathbf{a}', l; \lambda) g^{(n-1)}(\mathbf{b}', s; \lambda)=
g^{(n-k)}(\mathbf{a}'', l; \lambda) g^{(n-1)}(\mathbf{b}'', s; \lambda).
\end{equation}
Set $\lambda = (-1)^{n-k-1}l$ in the equation \eqref{eq:sep10_3}.
By applying 
item $(iv)$ of Proposition \ref{pr:nov27_1} on the right hand side, we 
see that the first $n-k$ columns of both sides of the equation
are contained in the $(n-k)$-dimensional subspace 
$\overline{\Theta}_{n-k} (\mathbf{a}')$, 
and also in $\overline{\Theta}_{n-k} (\mathbf{a}'')$ by \eqref{eq:nov27_2}.
On the other hand,
one observes that
the bottom left $(n-k) \times (n-k)$ submatrix of the
left hand side of the equation \eqref{eq:sep10_3} is independent of $\lambda$,
and its determinant is given by
the geometric coenergy function (Corollary 7.3 of \cite{F18}),
\begin{align}
E(\mathbf{b},\mathbf{a}) &= \Delta_{[k+1,n],[n-k]}(g^{(n-1)}(\mathbf{b}, s; \bullet) g^{(n-k)}(\mathbf{a}, l; \bullet)) \nonumber\\
&= \sum_{m=0}^{n-k} a^{(k,k)} a^{(k,k+1)} \cdots a^{(k,k+m-1)} 
b^{(1,k+m+1)}  b^{(1,k+m+2)}  \cdots b^{(1,n)}. \label{eq:nov13_7}
\end{align}
Since this quantity does not vanish, the first $n-k$ columns of 
both sides of equation \eqref{eq:sep10_3} always have full rank.
(This is true even if we set $s=l$ and $k$ is odd, hence the
matrix $g^{(n-1)}(\mathbf{b}, s; (-1)^{n-k-1}s)$ 
is not invertible and has rank $n-1$.)
Therefore we have $\overline{\Theta}_{n-k} (\mathbf{a}') =
\overline{\Theta}_{n-k} (\mathbf{a}'')$, and
then $\overline{\Theta}_{n-k} (\mathbf{b}') = \overline{\Theta}_{n-k} (\mathbf{b}'')$
by equation \eqref{eq:nov27_2} .
Since the map $\overline{\Theta}_{n-k}$
gives a bijection between $\overline{\mathbb{Y}}_k$
and ${\rm Gr}(n-k,n)_{>0}$, we have the desired result.
\qed

Based on this proposition, define
$R^{(s,l)}: \overline{\mathbb{Y}}_1 \times \overline{\mathbb{Y}}_k  \rightarrow  \overline{\mathbb{Y}}_k \times \overline{\mathbb{Y}}_1$ to be a rational map
given by $R^{(s,l)}:(\mathbf{b},\mathbf{a}) \mapsto (\mathbf{a}',\mathbf{b}')$.
This is the geometric $R$-matrix in the present case.

\subsubsection{Prerequisites for the Perron-Frobenius theorem.}\label{sec:3_2_2}
For any $\mathbf{x} \in \overline{\mathbb{Y}}_k$ let $\vec{P}(\mathbf{x})$ be an
${n \choose k}$-component vector defined by
\begin{equation}\label{eq:nov19_7}
\vec{P}(\mathbf{x}) = \left( \frac{P_I(\overline{\Theta}_{n-k} (\mathbf{x}))}{P_{[k+1,n]}(\overline{\Theta}_{n-k} (\mathbf{x}))} \right)_{I \in {[n] \choose n-k}},
\end{equation}
where the indices are assumed to be in lexicographic order
as in \eqref{eq:sep2_2}.
Then we see that its last element is always normalized to be one.

By using the formula \eqref{eq:sep8_1}
one can recover all
the elements of $\mathbf{x}= (x^{(i,j)})_{(i,j) \in R_k} \in \overline{\mathbb{Y}}_k$ from
the elements of the vector $\vec{P}(\mathbf{x})$.
Based on this fact, we can generalize Lemma \ref{lem:2} to the following:
\begin{lemma}\label{lem:sep2_3}
Let $R^{(s,l)}(\mathbf{b},\mathbf{a}) = (\mathbf{a}',\mathbf{b}')$.
Then the elements of $\mathbf{a}'\in \overline{\mathbb{Y}}_k$ are determined by the
following relation
\begin{equation}\label{eq:sep2_4}
\vec{P}(\mathbf{a}') = \frac{1}{E(\mathbf{b},\mathbf{a})} 
\mathsf{C}_{n-k}(g^{(n-1)}(\mathbf{b}, s; (-1)^{n-k-1} l) )\vec{P}(\mathbf{a}),
\end{equation}
where $\mathsf{C}_{n-k}$ denotes the $(n-k)$-th contravariant alternating tensor
representation \eqref{eq:sep2_2}, and $E(\mathbf{b},\mathbf{a})$ 
is the geometric coenergy function \eqref{eq:nov13_7}.
\end{lemma}
\proof
By Frieden's definition of the geometric $R$-matrix ((5.1) of \cite{F18}),
we have $\overline{\Theta}_{n-k} (\mathbf{a}')
= g^{(n-1)}(\mathbf{b}, s; (-1)^{n-k-1} l) \cdot \overline{\Theta}_{n-k} (\mathbf{a})$,
where the meaning of
$\cdot$ in the right hand side was given in section \ref{sec:1_3}.
As a matrix representative of $\overline{\Theta}_{n-k} (\mathbf{a}')$, 
define $\tilde{M}(\mathbf{a}')$ to be
an $n \times (n-k)$
matrix given by
$\tilde{M}(\mathbf{a}') = g^{(n-1)}(\mathbf{b}, s; (-1)^{n-k-1} l) M(\mathbf{a})$.
Then we have
\begin{equation}\label{eq:sep4_2}
\frac{P_{I}(\overline{\Theta}_{n-k} (\mathbf{a}'))}{P_{J}(\overline{\Theta}_{n-k} (\mathbf{a}'))}=
\frac{\Delta_{I, [n-k]}(\tilde{M} (\mathbf{a}'))}{\Delta_{J, [n-k]}
(\tilde{M} (\mathbf{a}'))},
\end{equation}
for any $(n-k)$-subsets $I, J$ of $[n]$.
On the other hand, by applying
the Cauchy-Binet formula to the definition of $\tilde{M} (\mathbf{a}')$ we have
\begin{align}
\Delta_{I, [n-k]}(\tilde{M} (\mathbf{a}')) &=
\sum_{J \in {[n] \choose n-k}} \Delta_{I,J} (g^{(n-1)}(\mathbf{b}, s; (-1)^{n-k-1} l) )
\Delta_{J,[n-k]} (M(\mathbf{a})) \nonumber\\
&=\Delta_{I,[n-k]}(g^{(n-1)}(\mathbf{b}, s; (-1)^{n-k-1} l) g^{(n-k)}(\mathbf{a}, l; \bullet)),\label{eq:sep4_1}
\end{align}
where we used the fact that the first $n-k$ columns of $g^{(n-k)}(\mathbf{a},l;\lambda)$
coincide with those of
the matrix $M (\mathbf{a})$.
Therefore by setting $I = [k+1,n]$ we have
$\Delta_{[k+1,n], [n-k]}(\tilde{M} (\mathbf{a}')) = E(\mathbf{b},\mathbf{a})$.
By using this identity
and the fact that the bottom left $(n-k) \times (n-k)$ submatrix of $M(\mathbf{a})$ is
upper uni-triangular, we can write the first equality of \eqref{eq:sep4_1} as
\begin{equation*}
\frac{\Delta_{I, [n-k]}(\tilde{M} (\mathbf{a}'))}{\Delta_{[k+1,n], [n-k]}(\tilde{M} (\mathbf{a}'))}=
\frac{1}{E(\mathbf{b},\mathbf{a})} 
\sum_{J \in {[n] \choose n-k}} \Delta_{I,J} (g^{(n-1)}(\mathbf{b}, s; (-1)^{n-k-1} l) )
\frac{\Delta_{J, [n-k]}(M(\mathbf{a}))}{\Delta_{[k+1,n], [n-k]}(M(\mathbf{a}))}.
\end{equation*}
By \eqref{eq:sep4_2}, this is the identity which we wanted to prove.
\qed

\begin{figure}[htbp]
\centering
\includegraphics[height=4cm]{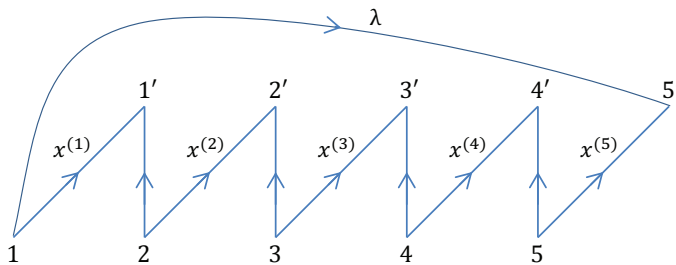}
\caption{An example of the network representation of the matrix $g(\mathbf{x}, s; \lambda) $ for $n=5$ case.
Vertical edges have weight 1.}
\label{fig:3}
\end{figure}

In what follows we omit the superscript of $g$ in the case of $k=1$,
hence $g(\mathbf{x}, s; \lambda)  = g^{(n-1)}(\mathbf{x}, s; \lambda) $
in agreement with the notation which we defined in \eqref{eq:sep29_1}.
Returning to the notations in the previous subsection,
let $(\mathbf{x}, s)$
denote an element of $\mathbb{Y}_1 =(\R_{>0})^{n-1} \times \R_{>0}$
with $\mathbf{x} = (x^{(1)},\dots,x^{(n-1)})$, and set $x^{(n)}:=s/(x^{(1)} \cdots x^{(n-1)})$.
Consider the matrix $\mathsf{C}_{r}(g(\mathbf{x}, s; \lambda) )$ given by \eqref{eq:sep2_2},
the $r$-th contravariant alternating tensor
representation of $g(\mathbf{x}, s; \lambda)$.
Note that the latter matrix has a network representation \cite{F19, F18}
as in Figure \ref{fig:3}.
Denote such a network by $\mathcal{N}_0$.

\begin{figure}[htbp]
\centering
\includegraphics[height=6cm]{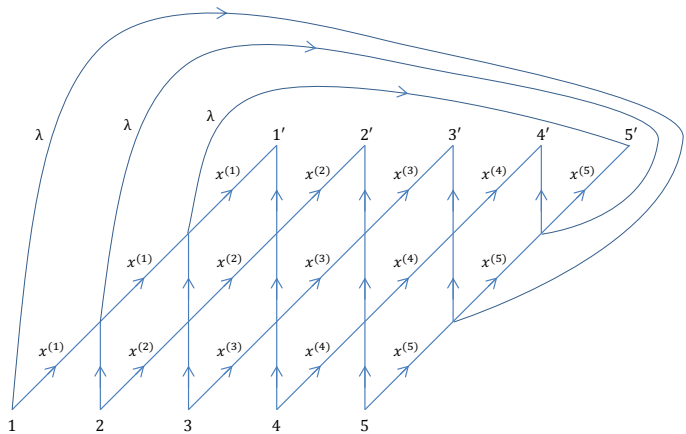}
\caption{A three times stack of the network representation of the matrix $g(\mathbf{x}, s; \lambda) $ for $n=5$ case.}
\label{fig:4}
\end{figure}

We say that a matrix is \textit{positive} if all its elements are positive real numbers.
In order to use the Perron-Frobenious theorem to define our system,
we need to prove:
\begin{proposition}\label{pr:sep15_1}
For any $\mathbf{x} \in (\R_{>0})^{n-1}$, $s,l \in \R_{>0}$ and $1 \leq r \leq n-1$,
some power of $\mathsf{C}_{r}(g(\mathbf{x}, s; (-1)^{r-1}l) )$ is
a positive matrix.
\end{proposition}
\proof
It suffices to show that there exists a positive integer $K$ such that
$\Delta_{I,J}(g(\mathbf{x}, s; (-1)^{r-1}l) ^K) > 0$ for any $I,J \in {[n] \choose r}$.
We begin with the fact that
the matrix $g(\mathbf{x}, s; \lambda)^K$ is 
represented by a planar network $\mathcal{N}$ that is given by 
stacking the network $\mathcal{N}_0$ for the matrix $g(\mathbf{x}, s; \lambda)$
up to $K$ times.
See Figure \ref{fig:4} for an example of $\mathcal{N}$ .
By the Lindstr$\ddot{\rm o}$m Lemma (See, for example Proposition 4.5 of \cite{F19}),
such a minor determinant is expressed as
\begin{equation}\label{eq:sep7_5}
\Delta_{I,J}(g(\mathbf{x}, s; (-1)^{r-1}l) ^K) =
\sum_{\mathcal{F} =(p_a; \sigma): I \rightarrow J}
{\rm sgn} (\sigma) {\rm wt} (\mathcal{F}),
\end{equation}
where the sum is over vertex-disjoint paths
(i.e.~no two of the paths share a vertex)
from $I = \{ i_1< \dots< i_r \}$ to 
$J = \{ j_1< \dots< j_r \}$ on the network $\mathcal{N}$ with $\lambda = (-1)^{r-1}l$,
and $\mathcal{F} =(p_a; \sigma)$ is a collection of $r$ paths $p_1, \dots, p_r$
such that $p_a$ starts at source $i_a$ and ends at sink $j'_{\sigma (a)}$,
for some permutation $\sigma \in S_r$.
The weight of $\mathcal{F}$ is defined as
$ {\rm wt} (\mathcal{F}) = \prod_{a=1}^r {\rm wt} (p_a)$,
where ${\rm wt} (p_a)$ is the weight of path $p_a$.

First we show that for sufficiently large $K$,
there exists at least one collection of vertex-disjoint paths 
from $I$ to $J$ for any $I,J \in {[n] \choose r}$.
\begin{enumerate}
\item
Suppose the elements of both $I$ and $J$ are consecutive mod $n$.
Let all the paths starting from the edges
labeled by $I$ go vertically upward on the
network.
Here we also regard the edges with weight $\lambda$ as
vertical ones.
Then they will arrive at the edges labeled by $J$ in a finite step,
if $K$ is sufficiently large.
\item
Suppose $I = I_1 \sqcup I_2$, where the elements of both $I_1$ and $I_2$ 
are consecutive mod $n$, and there is a gap between them.
Let all the paths
starting from the edges
labeled by $I_1$ go diagonally upward,
while those from the edges labeled by $I_2$ go vertically upward.
In a finite step, a \textit{merger} will occur.
That is, they will arrive at a collection of mod $n$ consecutive
edges on a common horizontal level.
\item
Suppose $I$ has more than 2 consecutive (mod $n$) subsets.
Let all the paths
starting from the edges
labeled by one of the subsets go diagonally upward,
while those from the edges labeled by the other subsets go vertically upward,
until a merger will occur.
Then by repeating this procedure we will come to the situation in (ii).
This together with the result in (i) is proving that for sufficiently large $K$
there exists at least one collection of vertex-disjoint paths 
from $I$ to $[r]$ for any $I \in {[n] \choose r}$.
\item
In the same way, we can show that for sufficiently large $K$,
there exists at least one collection of vertex-disjoint paths 
from $[r]$ to $J$ for any $J \in {[n] \choose r}$.
\item
By combining (iii) and (iv), we obtain the result which we wanted to show. 
\end{enumerate}

Now we show that for any collection of vertex-disjoint paths
$\mathcal{F} =(p_a; \sigma)$ from $I$ to $J$,
the summand ${\rm sgn} (\sigma) {\rm wt} (\mathcal{F})$
in the expression \eqref{eq:sep7_5}
is positive.
Because of the vertex-disjoint condition,
any $\sigma \in S_r$ must be either
a cyclic shift of the elements of $[r]$
or the identity map.
When going upward on the network
along the collection of paths $\mathcal{F}$, an elementary cyclic shift 
$\{ 1,2,\dots,r \} \rightarrow \{2,\dots,r,1\}$ may occur within one unit of the stack.
In that case, the ${\rm sgn} (\sigma)$ is multiplied by $(-1)^{r-1}$, while
an edge with weight $\lambda = (-1)^{r-1}l$ is picked up for the paths $\mathcal{F}$.
As a result, the summand ${\rm sgn} (\sigma) {\rm wt} (\mathcal{F})$
is always positive.
The proof is completed.
\qed

\vspace{5mm}
\par\noindent
Recall the definition of a Lax matrix $\mathcal{L}(| \mathbf{b} \rangle ; \lambda)$ in 
\eqref{eq:sep29_2}.
Then one easily sees that 
an obvious extension of Proposition \ref{pr:sep15_1} is the following:
\begin{corollary}
For any $| \mathbf{b} \rangle = (\mathbf{b}_1, \dots, \mathbf{b}_L) \in  (\R_{>0})^{L(n-1)}$, 
$s,l \in \R_{>0}$ and $1 \leq r \leq n-1$,
some power of 
$\mathsf{C}_{r}( \mathcal{L}(| \mathbf{b} \rangle ; (-1)^{r-1}l))$ is
a positive matrix.
\end{corollary}
By the Perron-Frobenius theorem, this corollary implies that
there exists a positive 
eivenvector 
of the matrix 
$\mathsf{C}_{r}( \mathcal{L}(| \mathbf{b} \rangle ; (-1)^{r-1}l))$
and it is unique up to a scalar multiple.

\subsubsection{Definition of the dynamical system.}\label{sec:3_2_3}
In order to define the time evolutions by carriers of ``rectangular tableaux'',
we first show the following:
\begin{lemma}\label{lem:sep10_1}
Let $A = \mathcal{L}(| \mathbf{b} \rangle ; (-1)^{n-k-1}l)$.
Then the
Perron-Frobenius eivenvector  $\vec{\xi}_{\rm PF}$ of
$\mathsf{C}_{n-k}( A )$ determines a unique point in
${\rm Gr}(n-k,n)_{>0}$.
\end{lemma}
\proof
Fix an arbitrary chosen $M_0 \in {\rm Gr}(n-k,n)_{>0}$. 
By using the same argument in the proof of Proposition \ref{pr:nov27_4},
we can define a sequence of $(n-k)$-dimensional subspaces $M_0, M_1, \dots  
\in {\rm Gr}(n-k,n)_{>0}$ 
recursively by the relation 
$M_i = A \cdot M_{i-1}$.
Then by using the Cauchy-Binet formula and
a consequence of the Perron-Frobenius theorem, we have
$\lim_{i \rightarrow \infty}(P_I(M_i))_{I \in {[n] \choose n-k}} = 
\lim_{i \rightarrow \infty} \mathsf{C}_{n-k}( A )^i (P_I(M_0))_{I \in {[n] \choose n-k}} = 
\vec{\xi}_{\rm PF}$ 
up to a scalar multiple. 
Since 
the closure (in the Hausdorff topology) 
of a totally positive Grassmannian ${\rm Gr}(n-k,n)_{>0}$
is
a totally nonnegative Grassmannian ${\rm Gr}(n-k,n)_{\geq 0}$ (Theorem 3.6 of \cite{L16}), 
this implies that there is a limiting point 
$M=\lim_{i \rightarrow \infty} M_i \in {\rm Gr}(n-k,n)_{\geq 0}$.
Then since $(P_I(M))_{I \in {[n] \choose n-k}}= 
\vec{\xi}_{\rm PF}$ is a positive vector,
this unique point $M$ actually lies in ${\rm Gr}(n-k,n)_{>0}$.
\qed

\begin{remark}
Since the Pl$\ddot{u}$cker coordinates of every point
in a Grassmannian must satisfy the Grassmann-Pl$\ddot{u}$cker relations
(See, for example Proposition 3.2 of \cite{F19}),
a positive vector with arbitrary chosen ${n \choose k}$ components does not
necessarily determines a point in ${\rm Gr}(n-k,n)_{>0}$.
\end{remark}

Recall that
$R^{(s,l)}: \overline{\mathbb{Y}}_1 \times \overline{\mathbb{Y}}_k  \rightarrow  \overline{\mathbb{Y}}_k \times \overline{\mathbb{Y}}_1$ is the rational map
given by the matrix equation \eqref{eq:sep10_3}.
As in the totally one-row tableaux case,
we define $R_i^{(s,l)}$ and $\mathcal{R}^{(s,l)}= R_1^{(s,l)} \circ \cdots \circ R_L^{(s,l)}$
to be maps which are now from $ (\overline{\mathbb{Y}}_1)^i \times \overline{\mathbb{Y}}_k
\times (\overline{\mathbb{Y}}_1)^{L-i} $ to 
$ (\overline{\mathbb{Y}}_1)^{i-1} \times \overline{\mathbb{Y}}_k
\times (\overline{\mathbb{Y}}_1)^{L-i+1}$, and from $ (\overline{\mathbb{Y}}_1)^L \times \overline{\mathbb{Y}}_k$ to 
$\overline{\mathbb{Y}}_k
\times (\overline{\mathbb{Y}}_1)^{L}$ respectively.
Given an arbitrary $(\mathbf{b}_1, \dots, \mathbf{b}_L, \mathbf{v}) \in  (\overline{\mathbb{Y}}_1)^L \times \overline{\mathbb{Y}}_k $,
let $\mathcal{R}^{(s,l)} (\mathbf{b}_1, \dots, \mathbf{b}_L, \mathbf{v}) = 
(\mathbf{v}_1, \mathbf{b}'_1, \dots, \mathbf{b}'_L)$.
It is depicted by the same diagram \eqref{eq:aug28_2}
but now $\mathbf{v}_i$'s are the elements of $\overline{\mathbb{Y}}_k = (\R_{>0})^{k(n-k)}$,
\begin{equation}\label{eq:aug28_2x}
\batten{\mathbf{v}_1}{\mathbf{b}_1}{\mathbf{b}_1'}{\mathbf{v}_2}\!\!\!
\batten{}{\mathbf{b}_2}{\mathbf{b}_2'}{\mathbf{v}_3}\!\!\!
\batten{}{}{}{\cdots\cdots}
\quad
\batten{}{}{}{\mathbf{v}_{L-1}}\,\,
\batten{}{\mathbf{b}_{L-1}}{\mathbf{b}_{L-1}'}{\mathbf{v}_{L}}\!\!\!
\batten{}{\mathbf{b}_L}{\mathbf{b}_L'}{\mathbf{v} \in \overline{\mathbb{Y}}_k.}
\end{equation}
Once again, 
by regarding the $\mathbf{b}_i$'s also as parameters, we consider an 
algebraic equation $\mathbf{v} = \mathbf{v}_1$ for the unknown $\mathbf{v}$
that assures the system of having a
periodic boundary condition.
Then we have:
\begin{proposition}\label{prop:sep10_2}
For any $s, l \in \R_{>0}$ and $| \mathbf{b} \rangle = (\mathbf{b}_1, \dots, \mathbf{b}_L) \in  (\overline{\mathbb{Y}}_1)^L$,
there is a unique positive real solution $\mathbf{v} \in \overline{\mathbb{Y}}_k = (\R_{>0})^{k(n-k)}$ 
to the equation $\mathbf{v} = \mathbf{v}_1$.
\end{proposition}
\proof
Let $A = \mathcal{L}(| \mathbf{b} \rangle ; (-1)^{n-k-1}l)$.
By repeated use of Lemma \ref{lem:sep2_3} we have
\begin{equation}\label{eq:oct2_1}
\mathsf{C}_{n-k}(A)\vec{P}(\mathbf{v}) = 
E(\mathbf{b}_{L},\mathbf{v})
E(\mathbf{b}_{L-1},\mathbf{v}_{L})
\cdots
E(\mathbf{b}_{1},\mathbf{v}_{2})
\vec{P}(\mathbf{v}_1),
\end{equation}
where $\mathbf{v}_i$'s are those determined by the diagram \eqref{eq:aug28_2x}.
\par\noindent
\textit{(Uniqueness.)}
Suppose there exist positive real solutions to the equation $\mathbf{v} = \mathbf{v}_1$.
From \eqref{eq:oct2_1} we see that
for any such solution $\mathbf{v} \in \overline{\mathbb{Y}}_k$ it is necessary for
$\vec{P}(\mathbf{v}) $ to be a positive eigenvector
of the matrix $\mathsf{C}_{n-k}(A)$.
By Lemma \ref{lem:sep10_1} 
there is a unique $M \in {\rm Gr}(n-k,n)_{>0}$ such that
$(P_I(M))_{I \in {[n] \choose n-k}}$ is the
Perron-Frobenius eivenvector of the matrix $\mathsf{C}_{n-k}(A)$. 
Therefore, if any such solution $\mathbf{v} \in \overline{\mathbb{Y}}_k$ exists, then it 
must be equal to the unique one
given from  
this $M$ by using the formula \eqref{eq:sep8_1},
because the map $\overline{\Theta}_{n-k}$
is a bijection between $\overline{\mathbb{Y}}_k$
and ${\rm Gr}(n-k,n)_{>0}$.
\par\noindent
\textit{(Existence.)}
Equation \eqref{eq:oct2_1} is valid for any $\mathbf{v} \in \overline{\mathbb{Y}}_k$
so in particular for the $\mathbf{v}$ obtained from the above mentioned unique
$M \in {\rm Gr}(n-k,n)_{>0}$.
On the other hand, for this $\mathbf{v}$ we also have
\begin{equation}\label{eq:nov27_5}
\mathsf{C}_{n-k}(A)\vec{P}(\mathbf{v}) =E^{(k)}_l \vec{P}(\mathbf{v}) 
\end{equation}
where $E^{(k)}_l$ is the dominant eigenvalue
of the matrix $\mathsf{C}_{n-k}(A)$.
By equating the right hand side of equation \eqref{eq:oct2_1} 
with that of \eqref{eq:nov27_5},
and noting that the last element of the vector $\vec{P}(\mathbf{x})$
defined in \eqref{eq:nov19_7} is always one for any 
$\mathbf{x} \in \overline{\mathbb{Y}}_k$,
we see that this $\mathbf{v}$ is indeed a solution to the algebraic equation 
$\mathbf{v} = \mathbf{v}_1$.
\qed

By combining the claims in Remark \ref{rem:nov19_3}
and Proposition \ref{prop:sep10_2}
we have the following:
\begin{theorem}\label{th:main2}
For any $s,l \in \R_{>0}$ and $(\mathbf{b}_1, \dots, \mathbf{b}_L) 
\in  (\R_{>0})^{L(n-1)} $,
there is a unique positive real solution 
$(\mathbf{v}, \mathbf{b}'_1, \dots, \mathbf{b}'_L) \in (\R_{>0})^{k(n-k)} \times (\R_{>0})^{L(n-1)}$ to the 
following matrix equation
\begin{equation}\label{eq:nov27_6}
g(\mathbf{b}_1, s; \lambda) \cdots g(\mathbf{b}_L, s; \lambda) g^{(n-k)}(\mathbf{v}, l; \lambda) =
g^{(n-k)}(\mathbf{v}, l; \lambda) g(\mathbf{b}'_1, s; \lambda) \cdots g(\mathbf{b}'_L, s; \lambda).
\end{equation}
\end{theorem}

Denote this unique positive real $\mathbf{v}$
by $\mathbf{v} = \mathbf{u}_l^{(k)} = 
\mathbf{u}_l^{(k)}(| \mathbf{b} \rangle) \in \overline{\mathbb{Y}}_k$.
Define $T^{(k)}_l: (\overline{\mathbb{Y}}_1)^L \rightarrow  (\overline{\mathbb{Y}}_1)^L$
to be a map given by
\begin{equation}\label{eq:sep10_4}
T^{(k)}_l (\mathbf{b}_1, \dots, \mathbf{b}_L) = (\mathbf{b}'_1, \dots, \mathbf{b}'_L),
\end{equation}
where the right hand side is determined by the relation
$\mathcal{R}^{(s,l)}  (\mathbf{b}_1, \dots, \mathbf{b}_L, \mathbf{u}_l^{(k)}) = 
(\mathbf{u}_l^{(k)}, \mathbf{b}'_1, \dots, \mathbf{b}'_L)$.
In the same way as in the totally one-row tableaux case,
we call this map a time evolution, and $\mathbf{u}_l^{(k)}= 
\mathbf{u}_l^{(k)}(| \mathbf{b} \rangle) $ a carrier for the state 
$| \mathbf{b} \rangle = (\mathbf{b}_1, \dots, \mathbf{b}_L)$ 
associated with $T^{(k)}_l$. 
This time evolution defines so far the most general case of the
closed geometric crystal chain.

Once again, any homogeneous state is a fixed point of this dynamical system.
To verify this claim, consider the result of
Proposition \ref{prop:sep10_2} for the one site $L=1$ case.
Then by using the explicit expression for the geometric $R$-matrix in \cite{F18}, 
we have $\mathbf{b}'_1 = \mathbf{b}_1$.
Then the same argument in the totally one-row tableaux case leads to
the required consequence.

By Theorem \ref{th:main2},
the time evolution \eqref{eq:sep10_4} is described by a Lax equation
for the matrix \eqref{eq:sep29_2} as
\begin{equation}\label{eq:oct22_1}
\mathcal{L}(T^{(k)}_l | \mathbf{b} \rangle ; \lambda)=
(g^{(n-k)}(\mathbf{u}_l^{(k)},l; \lambda))^{-1}
\mathcal{L}(| \mathbf{b} \rangle ; \lambda) g^{(n-k)}(\mathbf{u}_l^{(k)},l; \lambda).
\end{equation}
Therefore, the conservation laws considered in section \ref{sec:3_1_2}
are valid for any time evolutions $\{  T^{(k)}_l \}_{1 \leq k \leq n-1, l \in \R_{>0}}$.
Also,
since the geometric $R$-matrices satisfy the 
Yang-Baxter relation (Theorem 5.10(3) of \cite{F18}),
the same argument in section \ref{sec:2_2_1} assures the commutativity of
the time evolutions.
That is, we have $T^{(k_1)}_{l_1} \circ T^{(k_2)}_{l_2} = T^{(k_2)}_{l_2} \circ T^{(k_1)}_{l_1}$
for any $1 \leq k_1, k_2 \leq n-1$ and $l_1, l_2 \in \R_{>0}$.
\subsubsection{Invertibility.}\label{sec:3_2_4}
As in the $n=2$ case, every time evolution $T^{(k)}_l$ is invertible.
An explicit expression for the inverse map $(T^{(k)}_l)^{-1}$ is given as follows.
For any $n \times n$ matrix $X = (X_{ij})_{1 \leq i,j \leq n}$,
let ${\rm fl} (X)$ be an $n \times n$ matrix such that whose elements
in the $(i, j)$ position are given by ${\rm fl} (X)_{ij} = X_{n+1-j, n+1-i}$.
Then ${\rm fl}$ is an anti-automorphism of the ring of $n \times n$ matrices.

Define $S_l: \overline{\mathbb{Y}}_k \rightarrow \overline{\mathbb{Y}}_k$ 
to be a map given by $S_l(\mathbf{x}) = \mathbf{x}' = (x'^{(i,j)})_{(i,j) \in R_k}$
where $x'^{(i,j)}= x^{(k+1-i,n+1-j)}$ for $i+1 \leq j \leq i+n-k-1$, and
$x'^{(i,i)}= l/\prod_{m=k+1-i}^{n-i} x^{(k+1-i,m)}$,
for any $\mathbf{x} = (x^{(i,j)})_{(i,j) \in R_k} \in \overline{\mathbb{Y}}_k$.
This map is essentially identical with the geometric Sch$\ddot{\rm u}$tzenberger 
involution \cite{F19, F18}.
Then by the proof of Theorem 7.3 of \cite{F19} we have
\begin{equation}
{\rm fl} \circ g^{(n-k)}(\mathbf{x},l; \lambda) =
g^{(n-k)}(S_l(\mathbf{x}),l; \lambda).
\end{equation}
When $k=1$ we sometimes write 
$\mathbf{x} =  (x^{(1)},x^{(2)},\dots,x^{(n-1)}) \in \overline{\mathbb{Y}}_1$
and then we have $S_s(\mathbf{x}) =  (x^{(n)},x^{(n-1)},\dots,x^{(2)})$
where $x^{(n)}=s/(x^{(1)} \cdots x^{(n-1)})$.
By extending its definition,
for any $| \mathbf{b} \rangle = (\mathbf{b}_1, \dots, \mathbf{b}_L) \in  (\overline{\mathbb{Y}}_1)^L$ we let
$S_s | \mathbf{b} \rangle = (S_s(\mathbf{b}_L), \dots, S_s(\mathbf{b}_1))$.
Note that the order of the elements is reversed.
Since ${\rm fl}$ is an anti-automorphism we have
${\rm fl} (\mathcal{L}(| \mathbf{b} \rangle ; \lambda)) =
\mathcal{L}(S_s | \mathbf{b} \rangle ; \lambda).$
Therefore, by applying the anti-automorphism ${\rm fl}$ on both sides of \eqref{eq:oct22_1}
we obtain
\begin{equation*}
\mathcal{L}(S_s \circ T^{(k)}_l | \mathbf{b} \rangle ; \lambda)=
g^{(n-k)}(S_l(\mathbf{u}_l^{(k)}),l; \lambda)
\mathcal{L}(S_s| \mathbf{b} \rangle ; \lambda) (g^{(n-k)}(S_l(\mathbf{u}_l^{(k)}),l; \lambda))^{-1},
\end{equation*}
or equivalently
\begin{equation*}
\mathcal{L}(S_s| \mathbf{b} \rangle ; \lambda)=
(g^{(n-k)}(S_l(\mathbf{u}_l^{(k)}),l; \lambda))^{-1}
\mathcal{L}(S_s \circ T^{(k)}_l | \mathbf{b} \rangle ; \lambda) g^{(n-k)}(S_l(\mathbf{u}_l^{(k)}),l; \lambda).
\end{equation*}
By comparing with the definition of time evolution $T^{(k)}_l$,
we see from this equation that 
$S_s| \mathbf{b} \rangle = T^{(k)}_l \circ S_s \circ T^{(k)}_l | \mathbf{b} \rangle$.
Since $S_s$ is an involution and $| \mathbf{b} \rangle$ is an arbitrary 
element of $(\overline{\mathbb{Y}}_1)^L$, we have
\begin{equation}\label{eq:nov13_8}
(T^{(k)}_l)^{-1} = S_s \circ T^{(k)}_l \circ S_s.
\end{equation}
This is the explicit expression for the inverse map.

\subsubsection{Geometric lifting of the energy of paths.}\label{sec:3_2_5}
We reconsider the results on the conservation laws 
of the closed geometric crystal chains in section \ref{sec:3_1_2}.
In order to study the characteristic polynomial of the Lax matrix \eqref{eq:sep29_2},
we present some basic properties of the contravariant 
alternating tensor representation \eqref{eq:sep2_2}.
\begin{lemma}
For any $n \times n$ upper triangular matrix $B$,
the ${n \choose r} \times {n \choose r}$ matrix $\mathsf{C}_r (B)$
is also upper triangular.
\end{lemma}
\proof
Let $B = ( b_{ij} )_{1 \leq i,j \leq n}$ where 
$b_{ij} =0$ when $i >j$.
By definition, the matrix elements of $\mathsf{C}_r (B)$ are written as
\begin{equation}\label{eq:oct12_1}
\Delta_{I,J} (B) = \sum_{\sigma \in S_r} {\rm sgn} (\sigma)
b_{i_1, j_{\sigma (1)}} b_{i_2, j_{\sigma (2)}} \cdots b_{i_r, j_{\sigma (r)}},
\end{equation}
for $I = \{ i_1< i_2<\dots< i_{r} \}$ and $J = \{ j_1< j_2<\dots< j_{r} \}$.
We are to show that if $I > J$ in lexicographic order, then $\Delta_{I,J} (B) = 0$.
Let $q \in [r]$ be the smallest integer  
such that
$i_{q} > j_{q}$.
Choose an arbitrary permutation $\sigma \in S_r$.
If $\sigma (q) \leq q$ then $i_{q} > j_{q} \geq j_{\sigma (q)}$,
hence $b_{i_{q}, j_{\sigma (q)}}=0$.
Otherwise one has $\sigma (q) \geq  q+1$.
Then one of the $\sigma(q+1), \sigma(q+2), \dots, \sigma(r)$ is smaller than or equal to $q$.
Suppose $\sigma (a) \leq q$ for $ a \in [q+1, r]$.
Then $i_a > i_q > j_q \geq j_{\sigma (a)}$,
hence $b_{i_{a}, j_{\sigma (a)}}=0$.
Therefor, every term of the summation in \eqref{eq:oct12_1} is zero
when $I > J$. 
The proof is completed.
\qed

Since any square matrix is similar to some 
upper triangular matrix,
a consequence of this lemma is the following:
\begin{corollary}\label{cor:oct12_2}
Suppose a collection of complex numbers $\{ \mu_1, \ldots, \mu_n \}$ is the multiset \cite{Stanley97} of eigenvalues 
of an $n \times n$ matrix $A \in {\rm M}_n (\C)$, in which
their multiplicities are taken into account as repetitions of the elements.
Then the eigenvalues of the  ${n \choose r} \times {n \choose r}$ matrix 
$\mathsf{C}_r (A)  \in {\rm M}_{{n \choose r}} (\C)$ are given by the multiset
\begin{equation*}
\{ \mu_{i_1} \mu_{i_2} \cdots \mu_{i_r}  |
1 \leq i_1 < i_2 < \ldots < i_r \leq n \}.
\end{equation*}
\end{corollary}

Since the matrix elements of the Lax matrix \eqref{eq:sep29_2}
are polynomials of the loop parameter $\lambda$,
it is legitimate to substitute an arbitrary complex number into $\lambda$.
So in what follows we fix a $\lambda \in \C$.
Then the characteristic polynomial of the Lax matrix 
$\mathcal{L}(| \mathbf{b} \rangle ; \lambda) $
can be written as
\begin{equation*}
\det (x \mathbb{I}_n  - \mathcal{L}(| \mathbf{b} \rangle ; \lambda) )=
\prod_{i=1}^n (x - \mu_i ).
\end{equation*}
Since this polynomial is invariant under the time evolutions,
each eigenvalue $\mu_i $ (that depends on $\lambda \in \C$)
is a conserved quantity of the closed geometric crystal chain.
For any $I =\{ 1 \leq i_1 < \dots < i_{n-k} \leq n \} \in {[n] \choose n-k }$,
let $\mu_I = \mu_{i_1} \cdots \mu_{i_{n-k}}$.
By Corollary \ref{cor:oct12_2} we have
\begin{equation*}
\det \left( x \mathbb{I}_{{n \choose k}} -\mathsf{C}_{n-k}( \mathcal{L}(| \mathbf{b} \rangle ; \lambda ) \right)=
\prod_{I \in {[n] \choose n-k}} (x - \mu_I).
\end{equation*}
Therefore, the eigenvalues of the matrix $\mathsf{C}_{n-k}( \mathcal{L}(| \mathbf{b} \rangle ; \lambda ) $ are also conserved quantities.
In particular, the Perron-Frobenius eigenvalue of the 
\textit{monodromy matrix}
\begin{equation*}
\mathsf{M}_l^{(k)}(| \mathbf{b} \rangle):= 
\mathsf{C}_{n-k}( \mathcal{L}(| \mathbf{b} \rangle ; (-1)^{n-k-1}l)),
\end{equation*} 
is a positive real conserved quantity, which we denoted by $E_l^{(k)}$ 
in \eqref{eq:nov27_5}.
Then we have
\begin{equation}\label{eq:oct19_1}
\mathsf{M}_l^{(k)}(| \mathbf{b} \rangle)
\vec{P}(\mathbf{u}_l^{(k)}) = E_l^{(k)}\vec{P}(\mathbf{u}_l^{(k)}).
\end{equation}
By comparing this with the expression in \eqref{eq:oct2_1}
we have
\begin{equation*}
E_l^{(k)}
=E(\mathbf{b}_{L},\mathbf{u}_l^{(k)})
E(\mathbf{b}_{L-1},\mathbf{v}_{L})
\cdots
E(\mathbf{b}_{1},\mathbf{v}_{2}),
\end{equation*}
where $\mathbf{v}_i$'s are those determined by the diagram \eqref{eq:aug28_2x}
with $\mathbf{v}_1 = \mathbf{v}=\mathbf{u}_l^{(k)}$.
This expression with the explicit formula for the geometric coenegy
functions \eqref{eq:nov13_7} 
implies that the Perron-Frobenius eigenvalues
$E_l^{(k)}$ are the geometric liftings of the \textit{energy of paths},
a conserved quantity of
the (generalized periodic) box-ball systems \cite{FOY00} (See, also \cite{IKT12}).

\section{Summary and Discussions}\label{sec:4}
In this paper we proposed a method to
construct a new family of non-linear discrete integrable systems.
%
They are thought of as a geometric lifting of the 
integrable cellular automata known as the generalized periodic
box-ball systems.
By combining the G.~Frieden's work on the geometric $R$-matrix with
the Perron-Frobenious theorem, we were able to define a commuting family of
time evolutions $T^{(k)}_l$ for any $1 \leq k \leq n-1$ and $l \in \R_{>0}$
on the whole `phase space' $(\R_{>0})^{L(n-1)}$. 
In order to apply this theorem in linear algebra to our construction
of the non-linear integrable systems,
Lemmas \ref{lem:2} and \ref{lem:sep2_3} played an important role.
These lemmas claim that, although the geometric $R$-matrix is a non-linear rational map,
it can be viewed as almost a linear map provided that all its 
non-linearities have been pushed into a procedure of
the Gelfand-Tsetlin parametrization,
and also into a scalar factor called the geometric coenergy function.
Underlying ideas in these lemmas are inspired by 
a deep insight on
the property of this rational map (Remark 5.2 of \cite{F18}).

We have shown that
the time evolutions are described by Lax equations,
hence the conserved quantities are given by the coefficients of the
characteristic polynomial of the associated Lax matrices.
Equivalently, they are given as the eigenvalues of the Lax matrices,
and this fact gave us a simple
viewpoint that the dominant eigenvalue $E^{(k)}_l$
of the monodromy matrix is a geometric lifting of the energy of path (state)
of the corresponding generalized periodic box-ball system.
We noted this fact again here because
the energy of paths
is thought of as
one of the most important notions in the theory of 
integrable systems associated with crystals,
because it
is related not only to the number of solitons in the cellular automata, 
but also to the number of strings in the combinatorial Bethe ansatz for certain
integrable quantum spin chain models 
associated with such cellular automata \cite{KS08, KTT,KT10,KT2}. 

Regarding more about
the dominant eigenvalue $E_l^{(k)}$,
for any $k \in [n-1], l \in \R_{>0}$, and a given initial state,
it is a constant under any time evolutions. 
This fact allows us to adopt an interpretation that we can regard $E_l^{(k)}$ 
as one of the constant parameters of the system along with $s$ and $l$.
It also implies that for an actual realization of our dynamical system on 
a computer program,
though there is a possible purely numerical calculation process to
solve an algebraic equation for getting the dominant eigenvalue
of the monodromy matrix, we need to do that only once 
for each initial condition and
all the remaining calculation processes can be coded purely symbolic. 
In fact, under this interpretation,
the elements of the carrier $\mathbf{u}_l^{(k)}$
are viewed as rational 
(if not subtraction-free rational)
functions of $| \mathbf{b} \rangle$.
This is based on the fact that the vector
$\vec{P}(\mathbf{u}_l^{(k)})$ satisfies
the system of linear equation \eqref{eq:oct19_1}
with the constant parameters $s, l, E_l^{(k)}$,
and $\mathbf{u}_l^{(k)}$ is obtained from this vector
by using the rational map
\eqref{eq:sep8_1}.
To be more precise for the former claim, let $N={n \choose k}, 
\mathsf{M}_l^{(k)}(| \mathbf{b} \rangle) = (L_{i,j})_{1 \leq i,j \leq N}$,
and $\vec{P}(\mathbf{u}_l^{(k)})= (\mathcal{P}_1, \mathcal{P}_2, \dots, \mathcal{P}_{N-1}, 1)^t$.
Then we have
\begin{equation*}
\begin{pmatrix}
L_{11}-E_l^{(k)} & L_{12} & \dots & L_{1,N-1} \\
L_{21} & L_{22}-E_l^{(k)} & \dots & L_{2,N-1} \\
\vdots & \vdots & \ddots & \vdots \\
L_{N-1,1} & \dots & L_{N-1,N-2} &  L_{N-1,N-1}-E_l^{(k)} 
\end{pmatrix}
\begin{pmatrix}
\mathcal{P}_1 \\
\mathcal{P}_2 \\
\vdots \\
\mathcal{P}_{N-1}
\end{pmatrix}=
-\begin{pmatrix}
L_{1,N} \\
L_{2,N} \\
\vdots \\
L_{N-1,N}
\end{pmatrix}.
\end{equation*}
Therefore by the Cramer's rule the $\mathcal{P}_i$s are given by
rational functions of $L_{i,j}$s and $E_l^{(k)}$.
For instance, if $n=2$ the carrier $v (= \mathbf{u}_l^{(1)})$ in \eqref{eq:jul31_1}
is expressed as $v=(E_l^{(1)}-L_{22})/L_{21} = -L_{12}/(L_{11}-E_l^{(1)})$.

In the case of $n=2$, we conducted a detailed study on the 
system.
This includes an explicit list of the conserved quantities,
a discussion on the continuum limit of the system
to derive associated differential
equations, and also on the tropicalization of the system.
The latter study was done not only on 
the conserved quantities, but also on
the equation for the periodic boundary condition \eqref{eq:aug19_1}.
This result gives us an explanation
for the existence of the non-evolvable states in
the generalized periodic box-ball systems.
That is, while equation \eqref{eq:aug19_1} has a unique positive real solution,
its tropicalized counterpart \eqref{eq:aug20_1} does not always 
have a unique non-negative integer
solution.
We note once again that although the equation \eqref{eq:aug19_1} itself can be
tropicalized, its solution can not be tropicalized.
In this sense, we admit that
the closed geometric crystal chain is not literally a
geometric lifting of the generalized periodic box-ball system.

In the case of general $n$, we mainly restricted ourselves
to solve the problem of
whether we can define, if any, time evolutions compatible with
the periodic boundary condition. 
Now this work has been done, 
in future studies
we would like to clarify more detailed properties of
the system for the case of general $n$, 
such as taking a continuum limit of the system to obtain associated 
differential equations, and seeking an explicit formula
for the tropical limit of the energy of paths and
the equation for the periodic boundary condition,
as we have done in the case of $n=2$.
Also, there are many other remaining problems to be addressed;
whether we will be able to construct soliton solutions in our systems,
to give any explicit formulas to describe solutions to initial value problems,
to clarify the global structure of the iso-level sets of our dynamical systems
in the phase space $(\R_{>0})^{L(n-1)}$,
to define the action of the geometric crystal operators $e^c_i$'s
on the states of our systems,
and to clarify how the iso-level sets will be changed
under the operations of these operators.
Generalization of the states of the systems
from homogeneous paths associated with only one-row tableaux 
to inhomogeneous paths with more general rational rectangles,
may be another interesting problem.

Lastly, we give an additional 
discussion on the conserved quantities in the case of general $n$
in section \ref{sec:3_1_2}.
In fact, there is another way of giving them in terms of
polynomials of the variables $\{ b_i^{(j)} \}_{1 \leq i \leq L, 1 \leq j \leq n}$
with non-negative integer coefficients.
As we have seen in the proof of Proposition \ref{pr:sep15_1},
such a minor determinant is expressed as
\begin{equation*}
\Delta_{I,I}(\mathcal{L}(| \mathbf{b} \rangle ; \lambda)) =
\sum_{\mathcal{F} =(p_a; \sigma): I \rightarrow I}
{\rm sgn} (\sigma) {\rm wt} (\mathcal{F}),
\end{equation*}
where the sum is over vertex-disjoint paths from $I = \{ i_1< \dots< i_{n-k} \}$ to 
itself on a certain planar network associated with the matrix $\mathcal{L}(| \mathbf{b} \rangle ; \lambda)$,
and $\mathcal{F} =(p_a; \sigma)$ is a collection of $n-k$ paths $p_1, \dots, p_{n-k}$
such that $p_a$ starts at source $i_a$ and ends at sink $i_{\sigma (a)}$,
for some permutation $\sigma \in S_{n-k}$.
%
By a similar consideration in that proof, we see that for any $m$ the coefficient
of $\lambda^m$ in the polynomial 
${\rm Tr}\mathsf{C}_{n-k}( \mathcal{L}(| \mathbf{b} \rangle ; \lambda))$
is given by a polynomial of the variables $\{ b_i^{(j)} \}_{1 \leq i \leq L, 1 \leq j \leq n}$
with non-negative integer coefficients, multiplied by 
a sign factor $(-1)^{m(n-k-1)}$.
Therefore, by removing this sign factor, every conserved quantity contained in 
${\rm Tr}\mathsf{C}_{n-k}( \mathcal{L}(| \mathbf{b} \rangle ; \lambda))$
can be expressed by a polynomial
with non-negative integer coefficients.
This implies that, all the conserved quantities of the closed geometric crystal chains thus obtained
can be tropicalized.
As in the case of $n=2$, it is fairly reasonable to expect that
their tropicalization provides a collection of
piecewise-linear formulas for conserved quantities of the 
integrable cellular automata 
with periodic boundary conditions in \cite{KT10},
and their possible extensions from single box tableaux to
general one-row tableaux for the site variables in such cellular automata.

\vspace{5mm}
\noindent
{\it Acknowledgement}.
The authors thank Prof.~Takashi Arai and Prof.~Kazuo Hosomichi for valuable comments
on the second author's master thesis to which the present work
is partially related.

\vspace{0.5cm}

\end{document}